\documentclass[twocolumn,aps,prb,longbibliography,showpacs,10pt]{revtex4-2} 
\usepackage[normalem]{ulem}

\usepackage{amsfonts,amssymb,amsmath,graphicx,color}
\usepackage{array,multirow}
\usepackage{enumitem}
\usepackage{braket}
\usepackage[version=4]{mhchem}
\usepackage{float}
\usepackage[colorlinks=true,linkcolor=blue,citecolor=blue,urlcolor=blue]{hyperref}
\usepackage{hyperref}
\usepackage[caption=false]{subfig}
\usepackage{amsfonts}
\usepackage{amsmath}
\usepackage{commath}
\usepackage{amssymb}
\usepackage{placeins}
\usepackage{comment}
\usepackage{mathrsfs}

\def\ua{\uparrow}

\def\da{\downarrow}
\def\be{\begin{equation}}
\def\ee{\end{equation}}
\def\ber{\begin{eqnarray}}
\def\eer{\end{eqnarray}}

\newcommand{\ie}{{\it i.e.~}} 	
\newcommand{\eg}{{\it e.g.~}} 	
\newcommand\commentout[1]{}

\newcolumntype{P}[1]{>{\centering\arraybackslash}p{#1}}

\definecolor{greenPR}{rgb}{0.00, 0.6, 0.00}

\begin{document}

\title{Dirac Landau levels for surfaces with constant negative curvature}
\author{Maximilian Fürst$^1$}
\author{Denis Kochan$^{2,1,3}$}
\author{Ioachim-Gheorghe Dusa $^1$}
\author{Cosimo Gorini$^{4,1}$}
\author{Klaus Richter$^1$}
\affiliation{%
	$^1$Institut f\"ur Theoretische Physik, Universit\"at Regensburg, 93040 Regensburg, Germany\\
	$^2$Institute of Physics, Slovak Academy of Sciences, 84511 Bratislava, Slovakia \\
        $^3$Center for Quantum Frontiers of Research and Technology (QFort), National Cheng Kung University, Taiwan 70101, Taiwan \\
	$^4$SPEC, CEA, CNRS, Université Paris-Saclay, 91191 Gif-sur-Yvette, France 
}%
\date{\today}

\begin{abstract}
Studies of the formation of Landau levels based on the Schr\"{o}dinger equation for electrons constrained to curved surfaces have a long history. These include as prime examples surfaces with constant positive and negative curvature, the sphere~\cite{Haldane1983} and the pseudosphere~\cite{COMTET1987}. Now, topological insulators, hosting Dirac-type surface states, provide a unique platform to experimentally examine such quantum Hall physics in curved space. Hence, extending previous work we consider solutions of the Dirac equation for the pseudosphere for both, the case of an overall perpendicular magnetic field and a homogeneous coaxial, thereby locally varying, magnetic field. For both magnetic-field configurations, we provide analytical solutions for spectra and eigenstates.
For the experimentally relevant case of a coaxial magnetic field we find that the Landau levels split and one class shows a peculiar scaling $\propto B^{1/4}$, thereby characteristically differing from the usual linear $B$ and $B^{1/2}$ dependence of the planar Schr\"{o}dinger and Dirac case, respectively.
We compare our analytical findings to numerical results that we also extend to the case of the Minding surface.

\end{abstract}

\maketitle

\section{Introduction}

Topological classification of different states of matter rapidly evolved from a rare curiosity \cite{Volovik_Mineev_JETP1977,Volovik2007} into one of the hottest fields in physics \cite{Altland1997,Chiu2016,Schindler2018,Wu2019,Lieu2020}. Nowadays, topology serves as
an essential tool \cite{Volovik_UniverseInHeDroplet,HoravaPRL2008,Bzdusek2016,Bernevig2017} to characterize the ground state properties of interacting and non-interacting condensed matter-based systems -- comprising the quantum spin Hall Effect \cite{Volkov_Pankratov_1985,BHZ2006,Konig2007,Kane2005a}, topological insulators \cite{MooreBalents2007,Fu2007,Xia2009}, 3D semi-metals \cite{Wan2011a,Burkov2015,Wehling2014}, topological superconductors~\cite{SatoAndo2017}, to name a few.

On the other hand, there is also geometry entering into a local description of the dynamics. In the presence 
of a non-trivial metric or an emergent gauge field -- either in real or momentum space -- the underlying geometrical structure shows up in parallel transport.  Covariant derivatives thus enter the stage and play a decisive conceptual role. In single-particle non-relativistic quantum mechanics, the non-trivial geometry emerges when considering Bloch states of lattice-periodic Hamiltonians that give rise to a geometrical object known as the ``Fubini-Study-Berry connection''~\cite{Claassen2015PRL}. The latter is a silent eminence, for example, behind the popular family of various quantum Hall effects, including integer \cite{Klitzing1980}, spin \cite{Volkov_Pankratov_1985,BHZ2006,Konig2007,Kane2005a} and anomalous \cite{Nagaosa:RevModPhys2010} ones.
For concreteness, semiclassical $k$-space transport of a wave packet probes the presence of a local geometrical quantity reflected in the anomalous velocity \cite{Sundaram:PhysRevB1995}. The latter stems from non-trivial parallel transport itself, due to the emergent gauge field triggered by a coupling of the wave-packet center of mass momentum with underlying magnetic, spin-orbit or exchange-field-generated textures. Complementary, global quantities, {\it i.e.\ }objects integrated over $k$-space, such as conductances, polarizabilities or other response functions, reflect the underlying 
topological features, as first demonstrated by the celebrated TKNN formula \cite{TKKN1984}. The latter introduced concepts like Chern number and topological invariants into the field of condensed matter.

Apart from non-trivial geometrical features in momentum space, there are exciting possibilities for curved geometry in real space.  Positive spatial curvature is ``naturally realized in the laboratory'' on the surface of a sphere, for example, on a buckyball \cite{Terrones2003,Liang2016}.
Further options are offered by three-dimensional topological insulators (3DTIs) nanowires. The shape of the latter can be controlled at the nanoscale by growth \cite{Kessel2017,Behner2023} and etching techniques \cite{Ziegler2018}, so that the topological Dirac-like states existing on their surfaces propagate through curved two-dimensional space \cite{Kozlovsky2020,GRAF}. Indeed, when immersed in a magnetic field cylindrical TI nanowires with varying radius offer the possibility of combining non-trivial real space geometry with non-trivial reciprocal space one, characterizing e.g. the (integer) quantum Hall regime.  A notable consequence is that the nanowires' spectral and magnetotransport properties are governed by the subtle interplay between quantum confinement and geometry that involves Aharonov-Bohm (AB) and Berry phases \cite{Kozlovsky2020,GRAF,Xypakis-et-al-2020,Saxena2022}.

{Note that the curvature from such a shaping is smooth on the interatomic scale, and in particular does not arise from the distortion of the physical atomic lattice as may be the case \eg in graphene systems.  The curvature we consider rather emerges from the fact that topological states exist throughout the sample surface, and bends to follow the latter where this is etched or made to grow with a particular shape \footnote{Etching and growth clearly have material- and procedure-dependent effects on local atomic lattice properties, causing \eg local imperfections or surface reconstruction.  We neglect this physics throughout.}.

Crucially for our purposes, the shape of such 3DTI nanowire naturally leads to a (locally) negative spatial curvature.  Negative curvature is associated with interesting gravitational and cosmological phenomena, \eg \emph{event horizon} and \emph{Hawking and Unruh radiation}.  Its realization in condensed matter setups is challenging but possible, for example in certain meta-materials whose dielectric constant varies in a fashion effectively simulating the presence of a negatively curved metric \cite{Leonhardt2006,Genov2009,Bekenstein2017}.  It was thus possible to study analogues of the aforementioned gravitational and cosmological effects using ultra-short optical pulses \cite{Philbin2008}, acoustic waves \cite{Weinfurtner2011} and Bose–Einstein condensates \cite{Steinhauer2016}.  Effective negative curvature was also implemented via planar networks of superconducting microwave resonators \cite{Houck2012} which allow one 
to model strong interactions \cite{Anderson2016,Fitzpatrick2017,Kollar2019,Kollar2020} on the hyperbolic background. 

Generally, materials with nanoscale curved geometries possess many intriguing electronic and magnetic properties framing the fascinating subject of curved electronics \cite{Gentile2022}.
Following up on recent works by some of us \cite{Kozlovsky2020,GRAF}, we instead propose shaped 3DTI nanowires as a viable experimental platform \cite{Ziegler2018} to realize hyperbolic surfaces of (constant) negative curvature for Dirac-like electronic systems immersed in (strong) magnetic fields.

Landau levels have already been generally studied for surfaces with constant Gaussian curvature~\cite{Pnueli1994}. More specifically, those with a constant negative curvature in an overall perpendicular magnetic field includes the examination of Schrödinger-type~\cite{Ganeshan_2020, Grosche_1998, IKEDA1999, Grosche_1992, GROSCHE1988, Ferapontov_2001, Elstrodt1973} and of Dirac-type~\cite{Comtet:1984mm, GORBAR2008, Ludewig_2021} Landau levels on a hyperbolic plane, Schrödinger-type Landau levels on the Poincare disc \cite{Lisovyy_2007, COMTET1987} and Dirac-type Landau levels on a hyperboloid \cite{Demir_2020, Demir_20202}.

Here we  study the formation of Landau levels \footnote{Throughout the manuscript we will use the terms \emph{Landau Levels} and \emph{quantum Hall states} in a loose way to describe the eigenmodes of our system in a magnetic field, even if the field component orthogonal to the surface is non-homogeneous.  In this case such eigenmodes are not necessarily degenerate throughout the surface, and therefore not organized into globally-defined Landau Levels \cite{Kozlovsky2020, GRAF}.  Since the precise situation is described case-by-case, no confusion should arise.} on surfaces of revolution with constant negative Gauss curvature
 for two prominent geometries, the pseudosphere
and Minding surfaces, see Fig.~\ref{FIG_KAPPA}.
They both can be embedded in three-dimensional space and can even be potentially realized by means of TI nanowires with properly carved radial profiles. In view of such TI nanowires the Dirac case becomes relevant.
 We consider the solutions of the massless Dirac equation on the curved background and in non-trivial magnetic field configurations exploring geometrical effects stemming from the underlying spin and electromagnetic connections. 
In particular, we consider two configurations: (i) a magnetic field $\vec{B}_\perp$ (with constant field strength $B$) perpendicular to the surface of revolution; (ii) a coaxial magnetic field $\vec{B}_\parallel$ parallel to the rotation axis. The latter is much closer to experimental realizations, and to the best of our knowledge has not been addressed before~\footnote{In Ref.~\cite{Le_2019} very specific electromagnetic potentials have been addressed not including the two prominent field configurations mentioned. Note on the other hand that a similar problem on the standard sphere was recently considered \cite{Lee2009}.}.
For all geometries and field configurations considered, we calculate the corresponding eigenenergies and eigenfunctions, in most cases both analytically and numerically.

We find, among other results,  that Dirac Landau levels on the pseudosphere and the Minding surface in a strong coaxial $B$-field show a Zeeman-type splitting. Most notably,  these Landau levels exhibit a peculiar $B^{1/4}$-scaling that parametrically differs from the common $B^{1/2}$-dependence of the planar Dirac case, hence reflecting the effect of the Gaussian curvature in a measurable spectrum.

The paper is organized as follows: In Sec.~\ref{SecDiracCurved} we review the derivation of the massless Dirac equation for 
surfaces of revolution in external magnetic fields---taking into account concepts of spin connection and minimal coupling. 
This part involves elements from differential geometry.  It is included to make our presentation as self-contained as possible, but can be skipped by the reader not interested in such technical details.
In Sec.~\ref{SecDiracPseudo} we apply this approach
to surfaces of revolution with constant negative Gaussian curvature, namely, the pseudosphere and the Minding surface. We solve the underlying Dirac eigenvalue problems analytically in Sec.~\ref{Analytical}, and numerically in Sec.~\ref{Numerical}. Apart from this we also provide a detailed implementation scheme for practical lattice discretization of the underlying Dirac problems in curved spaces. We conclude in Sec.~\ref{SecConclusion}, and to provide a presentation  as self-contained as possible we include three technical appendices.

\begin{figure}[tbp]
{\centering
  \includegraphics[width=\columnwidth]{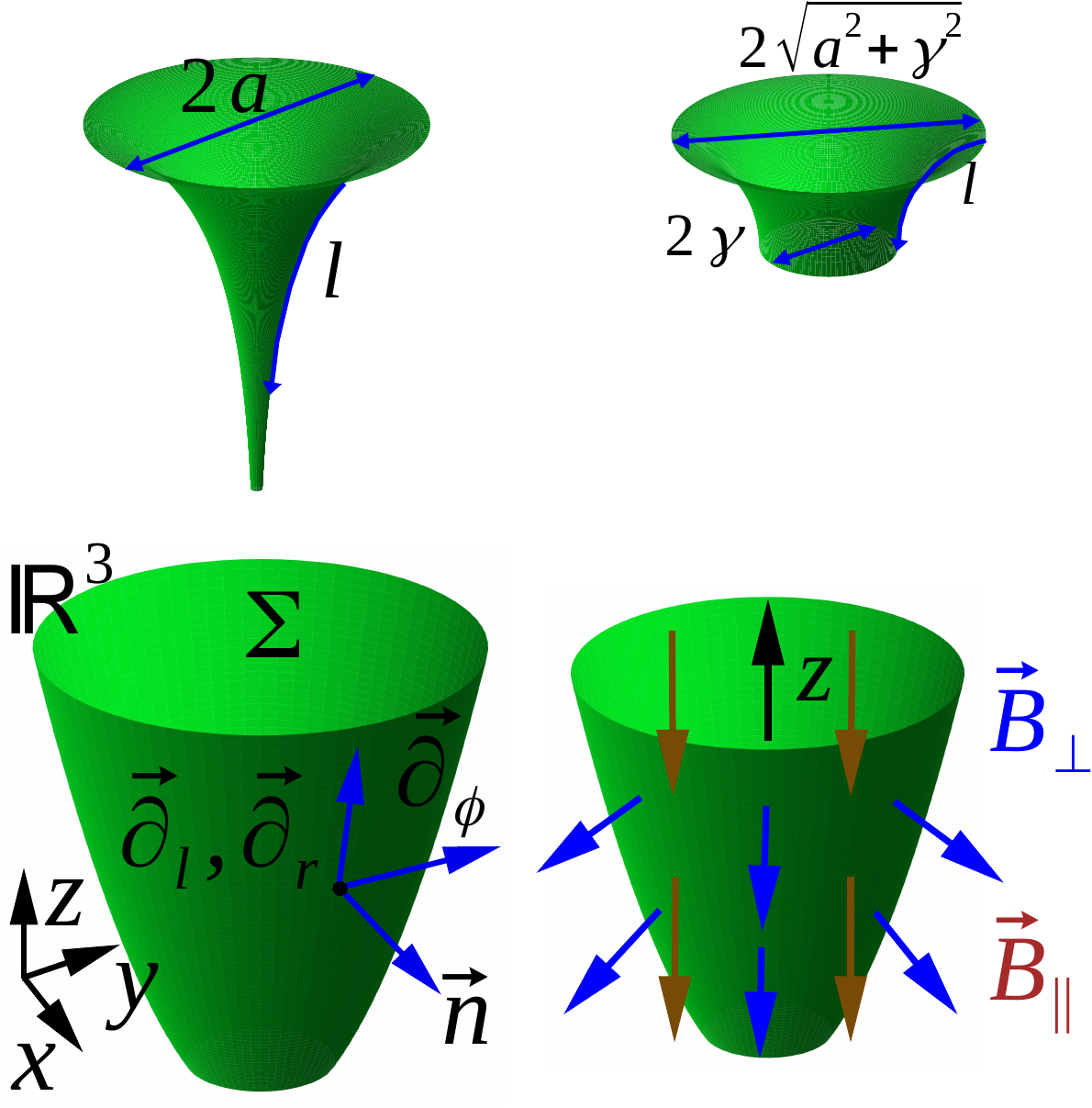}
  
}
  \caption{Surfaces of revolution embedded in $\mathbf{R}^3$: Pseudosphere (a) and Minding surface (b), including their angular and radial (arclength) parameterizations, the corresponding characteristic radii $r_{min}=\gamma$ and $r_{max}=\sqrt{a^2+\gamma^2}$ are defining outer and inner surface rims. For two magnetic configurations---the perpendicular magnetic field $\vec{B}_\perp$ points along the local outer 
  normal $\vec{n}\propto\vec{\partial}_\phi\times\vec{\partial}_r$ to the surface, while the coaxial field $\vec{B}_\parallel$ points along the rotational $z$-axis, we set both fields as positive
  if $\vec{n}\cdot\vec{B}_{\perp/\parallel}>0$,
  (c) Defining reference frames for a surface of revolution $\Sigma$ rotated along the 
  axial $z$-axis embedded in the Euclidean 3D space $\mathbf{R}^3$. Vectors 
  $\vec{\partial}_\phi$ and $\vec{\partial}_r\propto\vec{\partial}_l$ are tangential to 
  $\Sigma$.
  The perpendicular component of the magnetic field $\vec{B}$ points along the outer normal $\vec{n}\propto \vec{\partial}_\phi\times\vec{\partial}_r$,
  (d) Illustration of the two considered magnetic field configurations.
  }
  \label{FIG_KAPPA}
\end{figure}


\section{Dirac equation:
        from general formulation to curved 2D surfaces} \label{SecDiracCurved}

The formal construction of the Dirac operator for an arbitrary space-time manifold is a standard textbook material~\cite{D}. 
Therefore, for the sake of brevity, we just summarize the main conceptual steps and set up notation and convention. 
For a $(D+1)$-dimensional space-time manifold with a metric tensor $g_{\mu\nu}$ expressed in some 
local coordinates $\{x^\mu\}$ the corresponding Dirac equation for a massless particle reads~\cite{D,DIRAC_CURVED}
\begin{equation}
i \gamma^a e^\mu_a D_\mu \psi = 0 \quad ,
\end{equation}
where $e^\mu_a$ is the vielbein field, $g_{\mu \nu} = e_\mu^a e_\nu^b \eta_{ab}$ that brings the metric tensor $g_{\mu\nu}$ point-wisely into the canonical form $\eta_{ab}=\text{diag}{(-1,+1,\dots,+1)}$.
Correspondingly, $\gamma^a$ are the Dirac matrices with $\left\{\gamma^a, \gamma^b \right\} = 2 \eta^{a b}$. They  generate the underlying Dirac (Clifford) algebra, and 
\begin{equation}
D_\mu = \partial_\mu + \frac{1}{8}\hat{\omega}_{ab\mu}[\gamma^a,\gamma^b]
\end{equation}
stands for the covariant derivative
including the spin-connection~\footnote{The spin connection looks a priory like a mysterious 
object, while it couples spin to the curvature of the manifold, generally it ensures hermiticity of the Dirac operator with respect to a scalar product defined on the spinor fields. For 2D surfaces in Euclidean space the spin connection has a simple geometrical interpretation explained in Appendix \ref{APP_SPINCON}.}

\begin{equation}\label{Eq:SpinConnection}
\hat{\omega}_\mu^{ab} = e_\nu^a \Gamma_{\sigma \mu}^\nu e^{\sigma b} + e_\nu^a \partial_\mu e^{\nu b} \, ,
\end{equation}
see also Appendix~\ref{APP_SPINCON}. 

Throughout the paper we use Einstein's summation convention over repeated (one upper and one lower) Greek and Latin indices, while the metrics $g$ and $\eta$ raise and lower them, correspondingly. 
As usual, 
$\Gamma_{\sigma \mu}^\nu$ stand for the Christoffel symbols and $[{\cdot}\,,{\cdot}]$, $\{{\cdot}\,,{\cdot}\}$ for the associated commutators and anti-commutators, respectively. 
In presence of an additional electromagnetic $U(1)$-gauge field, described by the 1-form $A=A_\mu dx^\mu$ that couples to a charge $q$ of the massless particle, the above covariant derivative $D_\mu$ changes to $D_\mu-i\tfrac{q}{\hbar}A_\mu$ (minimal-coupling). Later, we consider a static magnetic field 
$\vec{B}$ and an electron with charge $q=-e<0$.

The scalar product of two spinor fields, $\psi_1$ and $\psi_2$, is defined as the volume integral over the D-dimensional spatial 
domain $\mathscr{D}$:
\begin{equation}\label{Eq:ScalarProductGeneral}
    \langle\psi_1|\psi_2\rangle=\int\limits_{\mathscr{D}}\underbrace{\sqrt{|\text{det}\,g_{\text{space}}|}\,dx^1\cdots dx^{\mathrm{D}}}_{\omega_g}\, (\overline{\psi}_1)^\mathcal{I} (\psi_{2})_\mathcal{I}\,,
\end{equation}
where the index $\mathcal{I}$ runs over the number of spinor-field components and the bar denotes 
complex conjugation. Furthermore, $\omega_g$ is an associated volume form on the spatial domain.

Particular examples of curved manifolds are 2D surfaces $\Sigma$ (and wave-guides) embedded in 3D Euclidean space, $\Sigma\subset\mathbf{R}^3$. Curved space-times corresponding to them emerge by building the Cartesian product of the 2D surface with the time-coordinate line $\mathbf{R}$, {\it i.e.}~$\mathbf{R} \times \Sigma$. 
In such a case, the curved space-time metric $g_{\mu\nu}$ is inherited from the flat pseudo-Euclidean metric $\langle{\cdot}\,,{\cdot}\rangle=\text{diag}(-v_F^2,+1,+1,+1)$ of the ambient space-time $\mathbf{R}\times\mathbf{R}^3$,
{\it i.e.}~$g_{\mu\nu}=\langle \vec{\partial}_\mu\bigl|_{\mathbf{R}\times\Sigma},\vec{\partial}_\nu\bigl|_{\mathbf{R}\times\Sigma}\rangle$. Here the speed of light gets substituted by an effective Fermi velocity $v_F$ as we intend to discuss fermions in metallic 2D systems. For practical computations we assume a value of $v_F = 5 \cdot 10^5$~m/s, which is a reasonable value considering topological insulators or Weyl-semimetals.
In the 2D case $\eta_{ab}=\text{diag}(-1,+1,+1)=e_a^\mu e_b^\nu g_{\mu\nu}$ -- with the time-like component $\eta_{00}=-1$, and the space-like components $\eta_{11}=\eta_{22}=+1$, $v_F$ becomes absorbed  into the definition of the vielbein fields $e_a^\mu$. Moreover, without loss of generality, we can choose the following representation of Dirac $\gamma$ matrices for the given $\eta_{ab}$:
\begin{equation}
\gamma^0 = i \sigma_z, \, \gamma^1 = \sigma_x, \, \gamma^2 = \sigma_y \quad ,
\end{equation}
where $\sigma_{x,y,z}$ are the standard $2\times 2$ Pauli matrices (for two spatial dimensions spinors are just two-component fields).
In what follows, we keep the discussion as simple as possible and just consider 2D surfaces that emerge as surfaces of revolution, say, with the axial axis chosen along the Cartesian $\hat{z}$ axis, see Fig.~\ref{FIG_KAPPA}.

The axial symmetry makes it advantageous to choose the azimuthal angle $\phi$ as one coordinate of the surface. There are many convenient choices for the remaining parameter, say, radius $r$ (if a surface is not a plain cylinder), the arclength $l$ when walking along the surface at a fixed azimuthal angle, or the altitude $z$ measured along the $z$-axis (if a surface is not flat).
It turns out that the radius $r$ offers a substantial advantage in analytical calculations, while the arclength $l$ is better for numerical lattice simulations, as discussed in Ref.~[\onlinecite{GRAF}]. Henceforth we assume $z(r)$ and $l(r)$ to be well-defined functions with differentiable inverses 
for $r$ ranging from \(r_{min}\) to a \(r_{max}\) -- meaning we exclude cylinders, pancakes and other surfaces containing them as parts, but consider those displayed in Fig.~\ref{FIG_KAPPA}. 
Changing parameterization from $r$ to $l$ is achieved by means of the relation (depending on whether $\tfrac{dl}{dr}\gtrless 0$):
\begin{equation}
    dl=\sqrt{dx^2+dy^2+dz^2}=\sqrt{1+\left(\tfrac{dz}{dr}\right)^2}\,|dr|=\pm\frac{dl}{dr}\,dr\,,
\end{equation}

For a rotationally symmetric spatial domain $\Sigma\subset\mathbf{R}^3$ parameterized by $(r,\phi)$ via the map
\begin{equation}
\Sigma: [r_{min}, r_{max}] \times [0, 2 \pi] \mapsto \mathbb{R}^3 
\ (r, \phi) \to
\begin{pmatrix}
	r \cos\phi\\
	r \sin\phi\\
	z(r)
\end{pmatrix} \, , 
\end{equation}
 the spatial tangent vectors to $\Sigma$ in the ambient Euclidean 3D space, see Fig.~\ref{FIG_KAPPA}, are spanned by
\begin{equation}\label{Eq:TangentVectors}
	 \vec{\partial}_{\phi} = \frac{\partial \Sigma}{\partial \phi}  =
	\begin{pmatrix}
		-r \sin(\phi)\\
		r \cos(\phi)\\
		0
	\end{pmatrix},
\ 
	\vec{\partial}_{r} = \frac{\partial \Sigma}{\partial r} =
	\begin{pmatrix}
		\cos\phi\\
		\sin\phi\\
		\frac{dz(r)}{dr}
	\end{pmatrix}.
\end{equation}
Then the space-time coordinates $\{x^\mu\}$ on $\mathbf{R}\times\Sigma$ read $(t,r,\phi)$. We define 
the orientation on $\mathbf{R}\times\Sigma$ by postulating $\{\vec{\partial}_\mu\}=\{\vec{\partial}_t,\vec{\partial}_\phi,\vec{\partial}_r\}$ to 
be the right-handed frame. Note that these vectors are orthogonal but not normalized in a sense 
of the ambient Pseudo-Euclidean space-time $\mathbf{R}\times\mathbf{R}^3$, with the flat metric 
$\langle \cdot, \cdot \rangle=\text{diag}(-v_F^2,+1,+1,+1)$.

The induced metric tensor $g_{\mu\nu}=\langle\vec{\partial}_\mu,\vec{\partial}_\nu\rangle$ on the curved
space-time $\mathbf{R}\times\Sigma$ reads
\begin{equation}
g_{\mu \nu} = \text{diag}\left(-v_F^2, r^2, 1 + \left(\tfrac{dz(r)}{dr}\right)^2\right)\,,
\end{equation}
while the vielbein field 
\begin{equation}
e_\mu^a =\text{diag}\left(v_F, r, \sqrt{1 + \left(\tfrac{dz(r)}{dr}\right)^2}\right)
\end{equation}
brings the metric tensor $g$ into the corresponding Sylvester normal form $\eta=\text{diag}(-1,+1,+1)$.
Note that the Greek indices $(t,\phi,r)$ stand for the chart-coordinates, with the corresponding coordinate vector fields 
$\vec{\partial}_t=\vec{\partial}_{0}, \vec{\partial}_\phi=\vec{\partial}_{1}, \vec{\partial}_r=\vec{\partial}_{2}$, 
while the Latin indices label their vielbein counterparts as captured by $e^\mu_a$.

Computing Christoffel's symbols \(\Gamma_{\mu \nu}^\sigma\), see Appendix \ref{APP_CHRISTOFFEL}, the spin connection $\hat{\omega}^{ab}_\mu$, Eq.~(\ref{Eq:SpinConnection}), yields only the non-zero entries
\begin{equation}\label{EQ_SPIN}
\hat{\omega}^{12}_\phi = - \hat{\omega}^{21}_\phi = \frac{1}{\sqrt{1 + \left(\frac{dz(r)}{dr}\right)^2}}=\frac{dr}{dl}\,.
\end{equation}
Taking all together and assuming a gauge field 1-form $A=A_\phi d\phi$, we obtain the following expressions for the covariant derivatives
of spinors on the curved space-time $\mathbf{R}\times\Sigma$:
\begin{align}
    D_0&=D_t=\partial_t \,,\label{Eq:D0}\\
    D_1&=D_\phi=\partial_\phi + \frac{i}{2} \hat{\omega}_\phi^{12} \sigma_z + i \frac{e}{\hbar} A_\phi\,,\label{Eq:D1}\\
    D_2&=D_r=\partial_r\,.\label{Eq:D2}
\end{align}
Now we can write down the final Dirac equation
\begin{equation}
i \gamma^a e^\mu_a D_\mu \psi =\left(-\sigma_z \frac{1}{v_F} \partial_t + i \sigma_y\, \hat{\omega}_\phi^{12}\, \partial_r + i \sigma_x \frac{1}{r} D_\phi \right) \psi = 0.
\end{equation}
 for massless charged fermions on $\mathbf{R}\times\Sigma$.
Note that by coincidence $\hat{\omega}_\phi^{12} = e_2^r$.

Making use of Dirac algebra and assuming stationary states, $\psi(t,r,\phi)=e^{-iEt/\hbar}\,\Psi(r,\phi)$, we obtain an eigenvalue problem for the spatial eigenspinors 
$\Psi(r,\phi)$ in terms of the equation
\begin{equation}\label{EQ_DIRAC_I}
\frac{\hat{H}}{\hbar v_F} \Psi \equiv
\left(i \sigma_x\, \hat{\omega}_\phi^{12}\, \partial_r - i \sigma_y \frac{1}{r} D_\phi \right) \Psi =
\frac{E}{\hbar v_F} \Psi 
\,.
\end{equation}

It remains to specify $A_\phi$ for the respective configuration of the magnetic field $\vec{B}$. 
We consider two cases: 

(i)~a magnetic field $\vec{B}_\perp$ with a constant magnitude perpendicular to the surface of revolution $\Sigma$
(in the sense of the ambient Euclidean space), 
and 

(ii)~ a homogeneous coaxial magnetic field $\vec{B}_\parallel$ being aligned in parallel to the axial $z$-axis of the 3D ambient space, i.e.~$\vec{B}_\parallel=(0,0,B)$.

The Faraday tensor $F$ at surface $\Sigma\subset\mathbf{R}^3$~\cite{D} for a generic magnetic field configuration $\vec{B}$ in $3D$ space takes the form 
\begin{equation}\label{EQ_VEC_OMEGA}
	F = -(\vec{B}\cdot\vec{n})\,\omega_g
\end{equation}
where $\omega_g$ is the ``volume-form'' defined on $\Sigma$ by the spatial part of the metric $g_{\mu\nu}$,
namely
\begin{equation}\label{EQ_VFORM}
\omega_g = \sqrt{\text{det}(g_{\Sigma})} \, d\phi \wedge dr = r \sqrt{1 + \left(\tfrac{dz(r)}{dr}\right)^2}  \, d\phi \wedge dr \, .
\end{equation}
Furthermore, $(\vec{B}\cdot\vec{n})$ stands for the Euclidean scalar product of the 3D vector $\vec{B}$ with the outer normal $\vec{n}$ to the surface $\Sigma$,
\begin{equation}
    \vec{n} = 
    \frac{\vec{\partial}_\phi \times \vec{\partial}_{r}}{||\vec{\partial}_\phi \times \vec{\partial}_{r}||} \! = \!
    \frac{1}{\sqrt{1+\left(\frac{dz}{dr}\right)^2}} \!
  \left(
  \tfrac{dz}{dr}\cos{\phi},\tfrac{dz}{dr}\sin{\phi},1
  \right)\, ,
  \label{eq:n}
\end{equation}
taken in the ambient space $\mathbf{R}^3$. In Eq.~(\ref{eq:n})  $\vec{\partial}_\phi$ and $\vec{\partial}_{r}$ are vectors in 
$\mathbf{R}^3$, see Eq.~(\ref{Eq:TangentVectors}), $\times$ is the usual vector product, and $||\cdot||$ denotes the Euclidean length of a 3D vector.

Throughout the text \(d \phi\) and \(d r\) are understood as differential 1-forms, and correspondingly $\omega_g$ and $F$, proportional to their wedge product $d\phi\wedge dr$, are 2-forms on $\Sigma$. Using 
the exterior derivative $d$, one can relate 
the 2-form $F$ to the electromagnetic potential 1-form $A$ \cite{D}:
\begin{equation}
dA = F = -(\vec{B}\cdot\vec{n})\,\omega_g\,.
\end{equation}
As both magnetic field configurations, $\vec{B}_\perp$ and $\vec{B}_\parallel$, are axially symmetric, and $\Sigma$ is a surface of revolution along the axial $z$-axis, $(\vec{B}\cdot\vec{n})$ only depends on the radial coordinate, {\it i.e.}~$(\vec{B}\cdot\vec{n})[r]$. Hence we use the symmetric electromagnetic gauge (what happens to be a Coulomb gauge)
$A =A_\phi(r) d\phi$, where the function $A_\phi(r)$ satisfies
\begin{equation}
dA=(\partial_r A_{\phi})\, dr\wedge d\phi\overset{!}{=}-(\vec{B}\cdot\vec{n})\,\omega_g \,.     
\end{equation}
Its solution can be expressed as an integral
\begin{eqnarray}\label{EQ_VEC}
A_\phi(r) &=& \int\limits^r (\vec{B}\cdot\vec{n})[r']\cdot r'\cdot\sqrt{1 + \left(\tfrac{dz(r')}{dr'}\right)^2} dr'\\
&=& \int\limits^{l} (\vec{B}\cdot\vec{n})[l']\cdot r(l') dl'\,. \notag
\end{eqnarray}
Explicit expressions for $A_{\phi,\perp}$ and $A_{\phi,\parallel}$ corresponding to $\vec{B}_\perp$ and $\vec{B}_\parallel$ are provided below.


\section{The Dirac equation on the pseudosphere and Minding surface} \label{SecDiracPseudo}

We apply the geometrical machinery reviewed in the previous section to surfaces of revolution 
that possess a constant negative Gaussian curvature 
\begin{equation}\label{EQ_Gauss}
\kappa=-\frac{1}{a^2} \, .
\end{equation}
Here $a>0$ -- the radius of curvature -- sets a natural spatial length scale of the problem, besides the (Fermi) wavelength.
A family of surfaces of revolution $\Sigma$ with a constant negative curvature $\kappa$, that are embeddable in $\mathbf{R}^3$,
$\Sigma:(r,\phi)\mapsto(r\cos\phi, r\sin\phi,z(r))$, 
are defined through the so-called \emph{Minding equation}~\cite{Popov-LobatchevskiGeom2014},
\begin{equation}\label{Eq:MindingDiffEq}
\left(\frac{dl}{dr}\right)^2\equiv 1+\left( \frac{dz(r)}{dr} \right)^2 = \frac{a^2}{r^2-\gamma^2}\, ,
\end{equation}
which needs to be solved for $z(r)$ (or $l(r)$) that enter the above parameterization of $\Sigma$. 
The Minding equation has two parameters controlling the ``shape'' of $\Sigma$: the radius of curvature $a>0$ (or curvature $\kappa= -1/a^2$ itself) and the parameter $0\leq\gamma<a$ that restrains ``an amount of the embeddable part'' with the ambient space $\mathbf{R}^3$.
The case $\gamma=0$ defines a {\em pseudosphere} or {\em Beltrami surface} -- a surface of revolution of the tractrix -- whereas the surfaces with $\gamma\neq 0$ are called {\em Minding surfaces}~\cite{Popov-LobatchevskiGeom2014}, see Fig.~\ref{FIG_KAPPA}.

While the Schrödinger-type Landau levels for the pseudosphere have been considered before \cite{GROSCHE1990,Can_Wiegmann_2017}, 
we will address in Secs.~\ref{Analytical} (analytics) and~\ref{Numerical} (numerics) the case of the Dirac equation for the pseudosphere and Minding surfaces.
The Dirac scenario acquires importance in view of possible experimental realizations in shaped 3DTI nanowires. Moreover, we will consider the experimentally relevant spatially homogeneous axial magnetic field, $\vec{B}_\parallel$, besides the conceptually simpler case of a constant perpendicular magnetic field, $\vec{B}_\perp$.

The embeddable part of a surface of revolution with a constant negative curvature within 3D Cartesian space only exists
for a certain range of radii. Since $( dz(r)/ dr)^2$ varies from zero to infinity, the middle-part of 
Eq.~(\ref{Eq:MindingDiffEq}) becomes equal or greater than 1. Consequently, the radii $r$ entering the right-hand side of Eq.~(\ref{Eq:MindingDiffEq}) should vary only 
within the interval $(r_{min}\!=\!\gamma,r_{max}\!=\!\sqrt{a^2+\gamma^2})$. 
\footnote{
Outside of this range, not considered here, the Minding surfaces can be treated as complex Riemann surfaces.}
As mentioned above, apart from the $r$ coordinate, also the arclength coordinate $l$ serves convenient for the parameterization of $\Sigma$.
Integrating the defining differential equation (\ref{Eq:MindingDiffEq}), one gets
\begin{eqnarray}\label{Eq:ArcLength}
l(r) &=& \pm\int_{r_0}^r \frac{a}{\sqrt{r'^2-\gamma^2}} dr'\,,
\end{eqnarray}
where the $\pm$ sign and the reference radius $r_0$ are free to choose. For the pseudosphere 
and Minding surface we set $l$ to zero at the rim with the largest radius, $r_{max}=\sqrt{a^2+\gamma^2}$, see Fig.~\ref{FIG_KAPPA}. Then the relation
\begin{align}\label{Eq:rVsl}
    l_{\gamma}(r)=a\ln{\left(\frac{a+\sqrt{a^2+\gamma^2}}{r+\sqrt{r^2-\gamma^2}}\right)}
\end{align}
holds equally for both cases $\gamma=0$ and $\gamma\neq 0$.\\

In what follows, we derive the eigenstates and eigenvalues of the Dirac Hamiltonians for massless particles moving on the pseudosphere and Minding surfaces subject to the magnetic field configurations, $\vec{B}_\perp$ and $\vec{B}_\parallel$, as specified above. 
Analytically, we mainly focus on eigenstates that are localized well inside the surface of revolution, i.e. their probability amplitudes are falling off sufficiently fast when approaching 
the edges. This sets the boundary conditions of our analytical spectral problem.
For the corresponding numerics, we represent a part of the surface by a discretized grid and diagonalize the Dirac Hamiltonians (with Wilson mass term~\cite{A}) using proper boundary conditions, see Sec.~\ref{Numerical}.

As $a>0$ sets the natural length scale of the problem, we rescale all quantities with a 
spatial dimension with $a$, using the tilde symbol, thus 
\begin{equation}
    \Tilde{r} = r/a \quad , \quad \Tilde{\gamma} = \gamma/a  \quad {\rm etc.}
\end{equation}
Using the general results from the previous section, and Eq.~(\ref{Eq:MindingDiffEq}) for $\sqrt{1+(dz(r)/dr)^2}$, we obtain the following expressions for the spin-connection, Eq.~(\ref{EQ_SPIN}), and for the corresponding electromagnetic gauge fields, Eq.~(\ref{EQ_VEC}):
\begin{eqnarray}
\hat{\omega}^{12}_\phi  = \sqrt{\tilde{r}^2-\tilde{\gamma}^2} \, ,
\end{eqnarray}
\begin{equation}
A_{\phi, \perp}  = B a^2 \sqrt{\tilde{r}^2-\tilde{\gamma}^2}\quad , \quad A_{\phi, \parallel} = B a^2 \frac{\Tilde{r}^2}{2} \, .
\end{equation}
Furthermore,
$\mathcal{E}=\hbar v_F/a$ sets the natural energy scale. Hence we normalize all quantities of dimension of energy by $\mathcal{E}$, but to avoid a proliferation of new symbols we do not introduce a new label for that, {\em i.e.}\ 
$H/\mathcal{E} \rightarrow H$. While for formulas and analytics we use the rescaled quantities, when plotting the spectra and effective potentials, and also for numerics we rather employ the dimensionful quantities and variables to grasp the magnitudes and units. In order to compare analytical and numerical results we use $v_F = 5 \cdot 10^5$~m/s and consider one representative pseudosphere ($\gamma=0$) and Minding surface ($\gamma=1$\,nm) with the same radius of curvature $a=60$\,nm.
This choice implies that, for magnetic fields with magnitude larger than $0.15$~T, the magnetic length $l_B = \sqrt{\hbar / e \abs{B}}$ becomes shorter than the considered curvature radius.

With these rescaling conventions the Dirac Hamiltonian, Eq.~\eqref{EQ_DIRAC_I}, as a function of $\gamma$ ($\gamma=0 \to$ pseudosphere, $\gamma\neq 0 \to$ Minding surface), reads
\begin{eqnarray}
\label{Eq:DiracForMinding}
\hat{H}_{\perp}
&=& 
i \sigma_x \sqrt{\Tilde{r}^2-\Tilde{\gamma}^2} \partial_{\Tilde{r}} -i \sigma_y \frac{1}{\Tilde{r}} \partial_\phi 
\notag\\
& & +\sigma_y \frac{\sqrt{\Tilde{r}^2-\Tilde{\gamma}^2}}{\Tilde{r}} \left(\frac{1}{\tilde{l}_B^2} + \frac{\sigma_z}{2} \right)\,,
\end{eqnarray}
\begin{eqnarray}
\label{Eq:DiracForMindingCoax}
\hat{H}_{\parallel}
&=& 
i \sigma_x \sqrt{\Tilde{r}^2-\Tilde{\gamma}^2} \partial_{\Tilde{r}} -i \sigma_y \frac{1}{\Tilde{r}} \partial_\phi \notag\\
& & +\sigma_y \left(\frac{\Tilde{r}}{2\tilde{l}_B^2} + \frac{\sqrt{\Tilde{r}^2-\Tilde{\gamma}^2}}{\Tilde{r}} \frac{\sigma_z}{2} \right)\, .
\end{eqnarray}
Here $H_\perp, H_\parallel$ stand for the Hamiltonian \eqref{EQ_DIRAC_I} respectively in the perpendicular and coaxial magnetic field configuration, while
$\tilde{l}_B^2$ is the dimensionless magnetic area multiplied by sgn$(B)$.
Specifically
\begin{equation}
    \frac{1}{\tilde{l}_B^2} = \text{sgn}(B)\,\frac{a^2}{l_B^2} \, ,
\end{equation}
where the magnetic length 
\begin{equation}
    l_B = \sqrt{\hbar / e \abs{B}}>0 \, .
\end{equation} 
These expressions hold for both magnetic field configurations, where $|B|$ stands for the magnitude of the $\vec{B}_{\perp/\parallel}$ field. 
We introduce the flux quantum $\Phi_0=h/2e$, the total area of the embeddable part of $\Sigma$, $\text{Area}_\Sigma=2\pi a^2$, as well as its projection onto the plane orthogonal to its symmetry axis, $\text{Area}'_\Sigma = \pi a^2 = \pi (r_{max}^2-r_{min}^2)$. One obtains for $\vec{B}_\perp$ and $\vec{B}_\parallel$ the perpendicular and parallel fluxes $|\Phi_\perp|=2\pi a^2 |B|$ and $|\Phi_\parallel|=\pi a^2 |B|$, correspondingly. 
Thus, all together
\begin{equation}\label{Eq:ell-in-flux}
    \frac{1}{\tilde{l}_B^2} = \text{sgn}(B)\,\frac{a^2}{l_B^2} = \text{sgn}(B)\frac{1}{\Phi_0}\times
    \begin{cases}
    \tfrac{1}{2}|\Phi_\perp|    & \text{for}\ \vec{B}_\perp\,,\\
    \ \ |\Phi_\parallel|  & \text{for}\ \vec{B}_\parallel\,.
    \end{cases}
\end{equation}
In what follows we keep the subscripts $\perp$ and $\parallel$ also for other quantities to distinguish and to trace their connections with the $\vec{B}_\perp$ and $\vec{B}_\parallel$ magnetic field configurations, respectively.

It is instructive to find discrete symmetries of $\hat{H}_{\perp/\parallel}$ to simplify the spectral analysis. The fact that $\sigma_z$ anticommutes with 
$\hat{H}_{\perp/\parallel}$ implies that the Hamiltonians have chiral symmetry:
\begin{equation}\label{EQ:symmetry1}
\sigma_z\, \hat{H}_{\perp/\parallel} = - \hat{H}_{\perp/\parallel}\,\sigma_z \, .
\end{equation}
As a consequence, the spectrum is symmetric with respect to \(E=0\), and if \(| \psi \rangle\) is an eigenstate of \(\hat{H}_{\perp/\parallel}\) with an eigenenergy \(E\), then 
\(|\Xi \rangle = \sigma_z | \psi \rangle\) is also an eigenstate with eigenvalue \(-E\). Positive and negative energy solutions are therefore directly linked by the chiral symmetry operator \(\sigma_z\).
Furthermore, having an eigenstate $|\psi \rangle$ of $\hat{H}_{\perp/\parallel}(\vec{B})$ with eigenenergy $E$ for a magnetic field $\vec{B}$, the state 
$|\chi \rangle=\sigma_y \mathscr{C} |\psi \rangle$ ($\mathscr{C}$ stands for the complex-conjugation operator) is an eigenstate of $\hat{H}_{\perp/\parallel}(-\vec{B})$ 
with the same energy, but for the opposite field configuration, {\it i.e.}\
\begin{equation}\label{EQ:symmetry2}
[\sigma_y\, \mathscr{C}]\, \hat{H}_{\perp/\parallel}(\vec{B}) 
= 
\hat{H}_{\perp/\parallel}(-\vec{B})\,[\sigma_y\, \mathscr{C}] \, .
\end{equation}
Thus, it is sufficient to examine only positively oriented magnetic 
fields, and for them to consider only eigenstates with non-negative energies.

\section{Analytical solutions}\label{Analytical}

\subsection{General considerations}
\label{subsec_general_considerations}

In the following we present analytical eigensolutions of the Hamiltonians given by 
Eqs.~(\ref{Eq:DiracForMinding}) and (\ref{Eq:DiracForMindingCoax}). For the pseudosphere we get full solutions for the perpendicular field, $\vec{B}_\perp$, and approximate solutions for the coaxial field, $\vec{B}_\parallel$,
while for the Minding surfaces we only find only the zero-energy eigenstates analytically in both magnetic configurations. 
Numerical solutions and a cross-check of analytic results with numerics will be presented in Sec.~\ref{Numerical}.

Since \(\hat{H}_{\perp/\parallel}\) in Eqs.~(\ref{Eq:DiracForMinding}) and (\ref{Eq:DiracForMindingCoax}) do not explicitly depend on \(\phi\), we use the separation ansatz
\begin{equation}\label{EQ_ANGULAR}
\Psi(\phi, \Tilde{r}) = \begin{pmatrix}
\Psi^\uparrow(\Tilde{r}) \\ \Psi^\downarrow(\Tilde{r})
\end{pmatrix} \, e^{i (m+\frac{1}{2}) \phi}, \, m \in \mathbb{Z}\,.
\end{equation}
Note that the boundary conditions of the angular part of the wavefunction are antiperiodic (see e.g.~discussion in Appendix~\ref{APP_SPINCON}, or Ref.~\cite{B}). This is in accordance with a local trivialization of the associated spin bundle that gives rise to the spin-connection in the form as presented by Eq.~(\ref{Eq:SpinConnection}).
For convenience we introduce effective potentials (in units of $\mathcal{E}=\hbar v_F/a$),
\begin{align}
V_{\perp} &= \frac{1}{\Tilde{r}} \left[\left(m+ \frac{1}{2}\right) + \frac{\sqrt{\Tilde{r}^2-\tilde{\gamma}^2}}{\tilde{l}_B^2} \right]\,,
\label{eq:V}\\
V_{\parallel} &= \frac{1}{\Tilde{r}}\left[\left(m+ \frac{1}{2}\right) + \frac{\Tilde{r}^2}{2\tilde{l}_B^2} \right]\,,
\label{eq:Vcoax}
\end{align}
and the auxiliary radial differential operators \footnote{These operators are adjoined to each other, apart from boundary terms stemming from integration by parts, assuming the spinor scalar product $\langle\psi_1|\psi_2\rangle$, Eq.~(\ref{Eq:ScalarProductGeneral}) that involves the volume form $\omega_g=\tfrac{r a}{\sqrt{r^2-\gamma^2}}d\phi\wedge d r$.}
\begin{equation}\label{Eq:AuxiliaryL}
    \hat{L}^{\pm}_{\perp/\parallel}=
    i \left[\sqrt{\tilde{r}^2-\tilde{\gamma}^2} \left(\partial_{\Tilde{r}} + \frac{1}{2 \Tilde{r}} \right) 
    \pm V_{\perp/\parallel}\right]\,.
\end{equation}
Using the separation ansatz, Eq.~(\ref{EQ_ANGULAR}), and $\mathcal{E}=\hbar v_F/a$ units the eigenproblem for the Dirac Hamiltonians, 
Eqs.~(\ref{Eq:DiracForMinding}) and (\ref{Eq:DiracForMindingCoax}), takes the form 
\begin{align}\label{Eq:RadialDirac}
&E 
\begin{pmatrix}
\Psi^\uparrow(\Tilde{r}) \\ \Psi^\downarrow(\Tilde{r})
\end{pmatrix} 
=
\begin{pmatrix}
0 & \hat{L}^{-}_{\perp/\parallel} \\ \hat{L}^{+}_{\perp/\parallel} & 0
\end{pmatrix} 
\begin{pmatrix}
\Psi^\uparrow(\Tilde{r}) \\ \Psi^\downarrow(\Tilde{r}) 
\end{pmatrix} \, .
\end{align}
The effective potentials $V_{\perp/\parallel}$ that depend on the angular quantum number $m$ 
act as variable mass term within the corresponding radial 
equations~\cite{Kozlovsky2020, GRAF, A}. 
The radii $\tilde{r}$ where $|V|$ becomes minimal indicate positions where the wavefunction amplitudes, and hence the corresponding probability densities get maximal.
Based on physical grounds, only minima which develop within the embeddable part of $\Sigma$ can accept electrons and thus contribute to the filling factor of the underlying Landau problem -- in full analogy with the formation of quantum Hall states in conventional flat samples. As stated before, we demand Landau levels to be localized within the surface, not at the boundary. Hence, we exclude those angular quantum numbers for which the minimum of $|V|$ is located at $\tilde{r} = 1$. Therefore by requiring $\tilde{r} \in \left[\tilde{\gamma}, \sqrt{1+\tilde{\gamma}^2}\right)$ one sets a natural cutoff for the critical value of $m$. 
For a positive (negative) field perpendicular to $\Sigma$, only negative (non-negative) angular quantum numbers are allowed:
\begin{eqnarray}
\mathcal{M}_{\perp} &=& \left\{
\begin{array}{cc}
    \left\{\left \lceil -\frac{\sqrt{1- \Tilde{\gamma}^2}}{\Tilde{l}_B^2} - \frac{1}{2} \right \rceil, ..., -1 \right \} \,, & 
    \ \ \text{for} \, B > 0 \\
    \\
    \left \{0, ..., \left \lfloor -\frac{\sqrt{1- \Tilde{\gamma}^2}}{\Tilde{l}_B^2} - \frac{1}{2} \right \rfloor \right \}, & \ \ \text{for} \, B < 0 
\end{array} \right. \label{Eq:mcritperp}
\end{eqnarray}
where $\lceil\cdot\rceil$ and $\lfloor\cdot\rfloor$ are ceiling and floor functions, respectively.

Contrary to this, for a coaxial field with any sign, negative and positive angular quantum numbers are possible:
\begin{eqnarray}
\mathcal{M}_{||} &=& \left \{\left \lceil -\frac{1}{2 l_B^2} - \frac{1}{2} \right \rceil, ..., \left \lfloor \frac{1}{2 l_B^2} - \frac{1}{2} \right \rfloor \right \} \, .
\label{Eq:mcritcoax}
\end{eqnarray}

We illustrate profiles of $|V_{\perp/\parallel}(\ell)|$ in the arclength coordinate, 
Eq.~(\ref{Eq:ArcLength}), for the pseudosphere in perpendicular, Fig.~\ref{FIG_VPOT}, and coaxial, Fig.~\ref{FIG_VPOT_COAX}, positive magnetic fields and different angular quantum numbers $m$. 
Obviously, minima of \(|V_{\perp}(l)|\) move from the outer rim ($l=0$) towards the cusp of the pseudosphere
when raising $B>0$ and lowering $|m|$. Based on what is said before, it can be easily seen that the minima within the surface form only for negative $m$ in the case magnetic field is perpendicular to the surface. In case of a coaxial field, both signs of $m$ are possible. However, it can clearly be observed that the potential wells for non-negative $m$ are higher in energy and show a different shape. Moreover, the confining potential wells become more pronounced upon increasing the magnetic field strengths. Contrary to this, in the absence of the field, the curves grow monotonously with $l$ and eigenstates only exist close to the outer rim.


\begin{figure}[tbp]
  \centering
  \includegraphics[width=1\columnwidth]{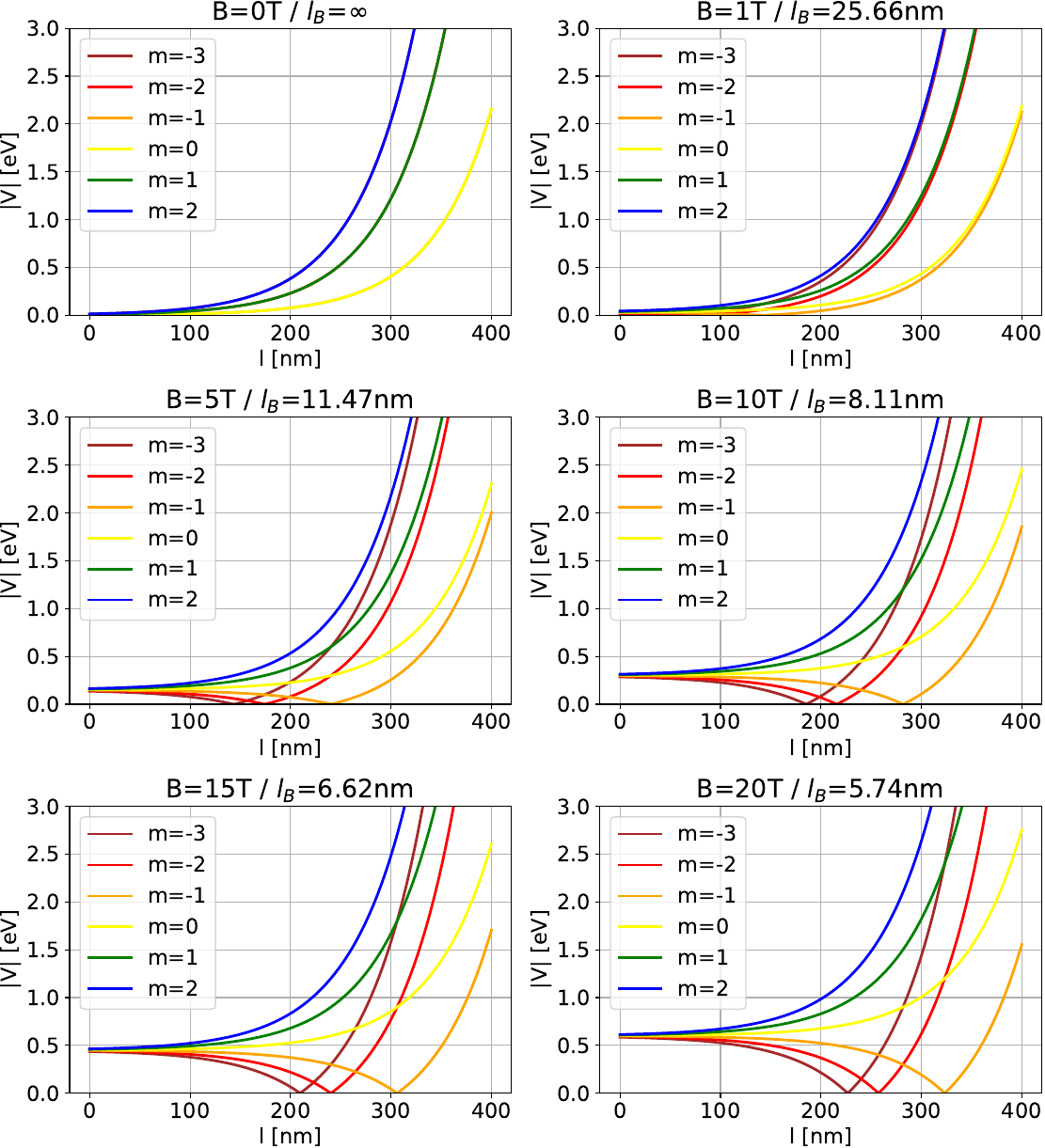}

  \caption{Effective potentials \(|V_{\perp}(l)|\), Eq.~(\ref{eq:V}), displayed as functions of the arclength coordinate \(l\) for a pseudosphere with \(a=60\,\)nm in a perpendicular magnetic field
  $\vec{B}_\perp$ pointing along the outer normal $\vec{n}$, see Fig.~\ref{FIG_KAPPA}(a).
  Different panels correspond to different strengths $B$, equivalently different magnetic lenghts $l_B$. 
  The colors code effective potentials for different angular quantum numbers \(m\). 
  For negative $m$, minima of \(|V_{\perp}(l)|\) move away from the outer rim of the pseudosphere ($l=0$) when
  raising $B$ and lowering $|m|$.
  }
  \label{FIG_VPOT}
\end{figure}
\begin{figure}[tbp]
  \centering
  \includegraphics[width=1\columnwidth]{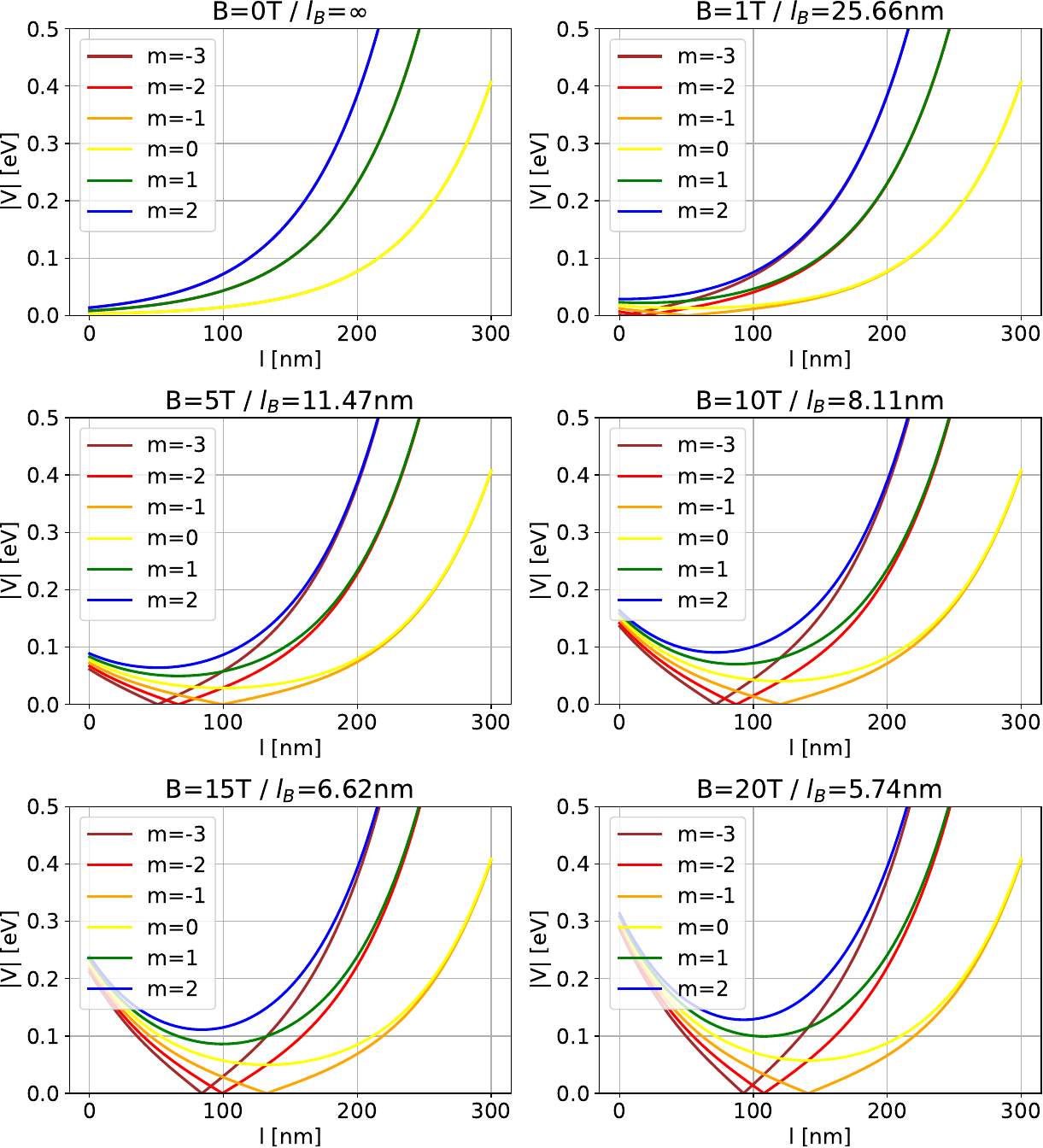}

  \caption{Effective potentials \(|V_{\parallel}(l)|\), Eq.~(\ref{eq:Vcoax}), displayed as functions of the arclength coordinate \(l\), for a pseudosphere with \(a=60\,\)nm in a coaxial magnetic field $\vec{B}_\parallel=(0,0,B)$, see Fig.~\ref{FIG_KAPPA}(a).
  Different panels correspond to different strengths $B$. The colors denote potentials for different angular quantum numbers \(m\). 
  Minima of \(|V_{\parallel}(l)|\) move away from the outer rim of the pseudosphere ($l=0$) when
  raising $B$ and lowering $|m|$.}
  \label{FIG_VPOT_COAX}
\end{figure}

To solve the radial equations, Eq.~(\ref{Eq:RadialDirac}), and compute
$\left( \Psi^\uparrow(\Tilde{r}),  \Psi^\downarrow(\Tilde{r}) \right)^\top$ for a general $E$, it is convenient to decouple the underlying system of first order differential equations,
\begin{align}
E\,\Psi^\uparrow(\Tilde{r}) &=\hat{L}^{-}_{\perp/\parallel}\,\Psi^\downarrow(\Tilde{r})\,,\label{Eq:FirstAndSecond1}\\
E\,\Psi^\downarrow(\Tilde{r}) &=\hat{L}^{+}_{\perp/\parallel}\,\Psi^\uparrow(\Tilde{r})\label{Eq:FirstAndSecond2}\,, 
\end{align}
by plugging one into the other. This leads to two separate second order differential equations for the two spinor components at eigenenergy $E$ (in units of $\mathcal{E}=\hbar v_F/a$):
\begin{align}
E^2\,\Psi^\uparrow(\Tilde{r}) &=\hat{L}_{\perp/\parallel}^{-}\hat{L}_{\perp/\parallel}^{+}\Psi^\uparrow(\Tilde{r})\,,\label{Eq:FirstAndSecond3}\\
E^2\,\Psi^\downarrow(\Tilde{r}) &=\hat{L}_{\perp/\parallel}^{+}\hat{L}_{\perp/\parallel}^{-}\Psi^\downarrow(\Tilde{r}) \, ,
\label{Eq:FirstAndSecond4}
\end{align}
that we consider below.


\subsection{Pseudosphere}
\subsubsection{Zeroth Landau level}

For $E=0$, the system of first order differential equations, Eqs.~(\ref{Eq:FirstAndSecond1})~and~(\ref{Eq:FirstAndSecond2}), decouples and one needs to solve 
\begin{equation}
    \hat{L}^{+/-}_{\perp/\parallel}\,\Psi^{\uparrow/\downarrow}(\Tilde{r})=0\,.
    \label{eq:Lpm}
\end{equation}
In the following, the solutions of the equations containing the superscript $\pm$ are marked with the same symbol.
Substituting $\tilde{\gamma}=0$ into the expressions for $V_{\perp/\parallel}$ and $\hat{L}^{\pm}_{\perp/\parallel}$ leads 
to simple radial equations with straightforward solutions.
The full spinorial solutions $\bigl( \Psi^\uparrow(\Tilde{r},\phi),\Psi^\downarrow(\Tilde{r},\phi) \bigr)^\top$ for the zeroth Landau level ($n=0$) read:
\begin{align}
\Psi_{0, \perp}^+(\phi,\Tilde{r}) 
&= 
\Tilde{r}^{-\frac{1}{2}-\frac{1}{\tilde{l}_B^2}} e^{+\frac{1}{\Tilde{r}} \left(m+\frac{1}{2}\right)+i \left(m+\tfrac{1}{2} \right)\phi}
\begin{pmatrix}
1 \\
0
\end{pmatrix} \,,\label{Eq:PseudoSpehereZeroLandauLevelUp}\\
\Psi_{0, \perp}^-(\phi,\Tilde{r})
&= \Tilde{r}^{-\frac{1}{2}+\frac{1}{\tilde{l}_B^2}} e^{-\frac{1}{\Tilde{r}} \left(m+\frac{1}{2}\right)+i \left(m+\tfrac{1}{2} \right)\phi}
\begin{pmatrix}
0 \\
1
\end{pmatrix} \,.\label{Eq:PseudoSpehereZeroLandauLevelDown}
\end{align}

\begin{align}
\Psi_{0, \parallel}^+(\phi,\Tilde{r}) 
&= 
\Tilde{r}^{-\frac{1}{2}} e^{-\frac{\Tilde{r}}{2\Tilde{l}_B^2}} e^{+\frac{1}{\Tilde{r}} \left(m+\frac{1}{2}\right)+i \left(m+\tfrac{1}{2} \right)\phi}
\begin{pmatrix}
1 \\
0
\end{pmatrix} \,,\label{Eq:PseudoSpehereZeroLandauLevelUpCoax}\\
\Psi_{0, \parallel}^-(\phi,\Tilde{r})
&= \Tilde{r}^{-\frac{1}{2}} e^{\frac{\Tilde{r}}{2\Tilde{l}_B^2}} e^{-\frac{1}{\Tilde{r}} \left(m+\frac{1}{2}\right)+i \left(m+\tfrac{1}{2} \right)\phi}
\begin{pmatrix}
0 \\
1
\end{pmatrix} \,.\label{Eq:PseudoSpehereZeroLandauLevelDownCoax}
\end{align}
To check which of them are physical we demand the amplitude of the wavefunction to vanish when $\Tilde{r} \to 1$ and $\Tilde{r} \to 0$.

For positive fields, solutions $\Psi^+_{0,\perp/\parallel}$ with negative $m$ are physically relevant, while for the negative ones, their counterparts $\Psi^-_{0,\perp/\parallel}$ with non-negative $m$ span the zeroth Landau level. Also for the coaxial field, zero energy states only exist for one sign of the angular quantum number $m$. This is in accordance with our analysis of the effective potentials, since they are located higher in energy for one branch of $m$. In any case, for a non-zero magnetic field $\vec{B}_{\perp/\parallel}$ the 
\textit{zeroth Dirac Landau level becomes fully spin polarized with spins pointing along} $\text{sgn}(B)\,\vec{n}$. A similar analysis was done for a 2D sphere in a homogeneous perpendicular field, see e.g.~\cite{Schliemann2008, Lee2009, SPHERE}, where the zeroth Dirac Landau level is also spin polarized but electron spins are aligned along $-\text{sgn}(B)\,\vec{n}$.

Fig.~\ref{FIG_WF0} shows the probability densities of representative zero-energy 
eigenstates in the respective effective potentials for $B_\perp$; results for $B_\parallel$ 
look very similar.

\begin{figure}[tbp]
  \centering
  \includegraphics[width=1\columnwidth]{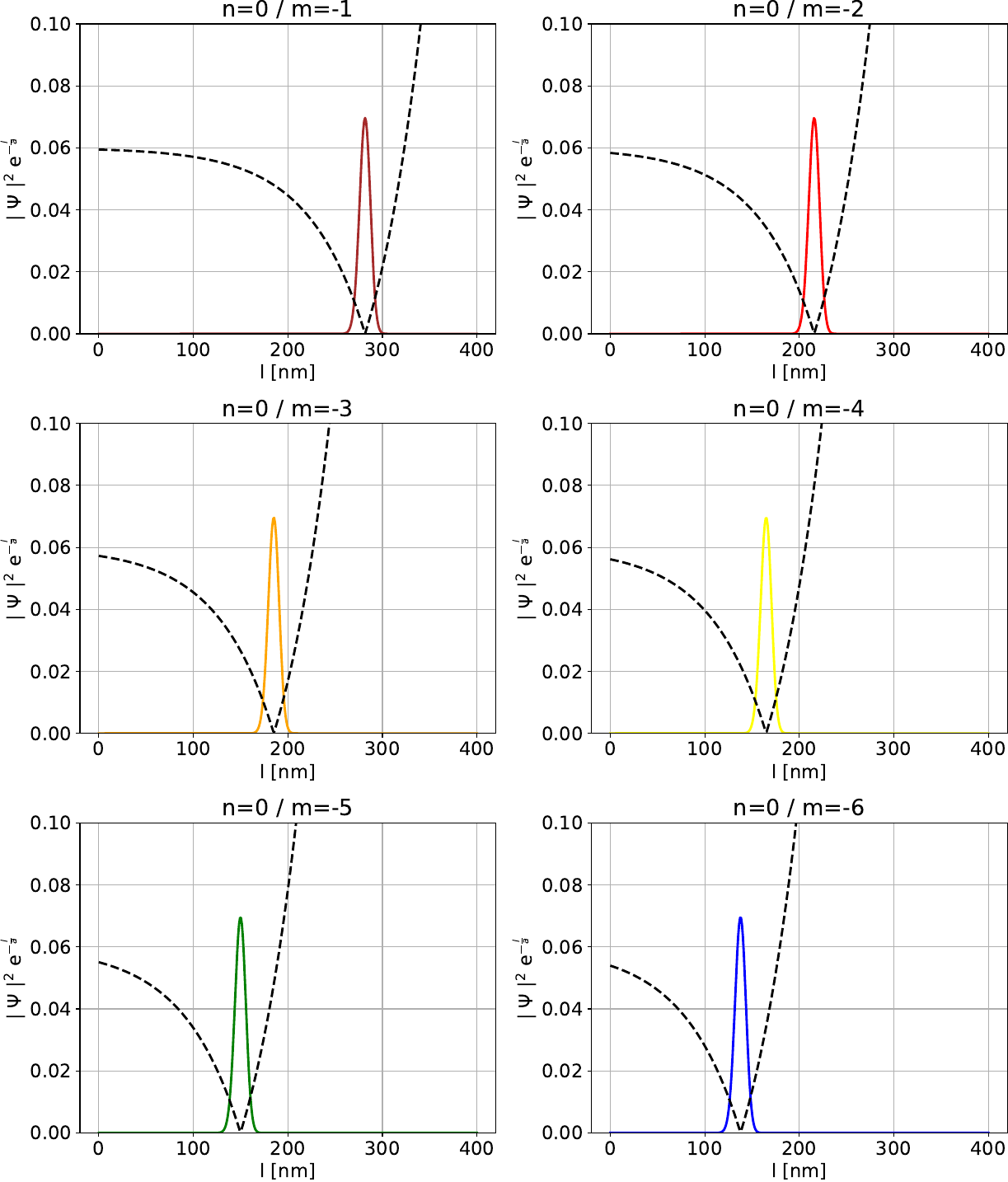}

  \caption{Pseudosphere radial probability densities \(\abs{\Psi}^2 e^{-\frac{l}{a}}\) (solid lines, unnormalized) for selected zero-energy eigenstates of $\hat{H}_{\perp}$, 
  Eq.~(\ref{Eq:DiracForMinding}), for different angular quantum numbers $m$.
Probability density maxima of these zeroth Landau levels ($n=0$) are localized at the minima of the corresponding effective potentials 
  \(\abs{V_\perp(l)}\) (dashed lines).
  Here: radius of curvature of pseudosphere \(a=60\)\,nm and perpendicular magnetic 
  field \(B=10\)\,T.}
  \label{FIG_WF0}
\end{figure}

\subsubsection{Higher Landau levels: general}

To simplify the analysis, for higher Landau levels we consider just positive magnetic fields, $\tilde{l}_B^2>0$, as the negative ones can be obtained by means of the $\sigma_y \mathscr{C}$ symmetry stipulated by
Eq.~(\ref{EQ:symmetry2}).
We start from the second order differential equations, Eq.~\eqref{Eq:FirstAndSecond3}, for the up component,
\begin{eqnarray}\label{EQ_+}
0 &=& \left[\Tilde{r}^2 \partial_{\Tilde{r}}^2 + 2 \Tilde{r} \partial_{\Tilde{r}} + \tilde{r} \frac{\partial V_{\perp/\parallel}}{\partial \Tilde{r}} - V_{\perp/\parallel}^2 + \frac{1}{4} + E^2 \right] \Psi^{\uparrow}(\Tilde{r}) \,, \notag \\
&&
\end{eqnarray}

and Eq.~\eqref{Eq:FirstAndSecond4}, for the down component,
\begin{eqnarray}\label{EQ_-}
0 &=& \left[\Tilde{r}^2 \partial_{\Tilde{r}}^2 + 2 \Tilde{r} \partial_{\Tilde{r}} - \tilde{r} \frac{\partial V_{\perp/\parallel}}{\partial \Tilde{r}} - V_{\perp/\parallel}^2 + \frac{1}{4} + E^2 \right] \Psi^{\downarrow}(\Tilde{r}) \,, \notag \\
&&
\end{eqnarray}
For negative $m$, We assume that the excited states show the same asymptotic behaviour of the radial part as the zero energy solution $\Psi_{0, \perp/\parallel}^{+}(\phi, \Tilde{r})$, Eq.~(\ref{Eq:PseudoSpehereZeroLandauLevelUp}). It turns out to be enough to only consider the equation for the up component, Eq.~\eqref{EQ_+}. This implies the following ansatz for the perpendicular and coaxial field, respectively:
\begin{equation}\label{Eq:AnsatzForHigherLL}
\Psi_{\perp}^{\uparrow}(\Tilde{r}) = p_{\perp}^{\uparrow}(\Tilde{r}) \, \Tilde{r}^{-\frac{1}{2}-\frac{1}{\tilde{l}_B^2}} \, e^{\frac{1}{\Tilde{r}} \left(m+\frac{1}{2}\right)} \,,
\end{equation}
\begin{equation}\label{Eq:AnsatzForHigherLLcoax}
\Psi_{\parallel}^{\uparrow}(\Tilde{r}) = p_{\parallel}^{\uparrow}(\Tilde{r}) \, \Tilde{r}^{-\frac{1}{2}} \, e^{-\frac{\tilde{r}}{2 \Tilde{l}_B^2}} \, e^{\frac{1}{\Tilde{r}} \left(m+\frac{1}{2}\right)} \,,
\end{equation}
where \(p_{\perp/\parallel}^{\uparrow}(\Tilde{r})\) are so far unknown functions.
Inserting the above ansatz into Eq.~\eqref{EQ_+} gives the following differential equations for \(p_{\perp/\parallel}^{\uparrow}(\Tilde{r})\):
\begin{eqnarray} \label{Eq:AnsatzForHigherLLPolynomial}
&&\Tilde{r}^2 \frac{d^2 p^\uparrow_{\perp}}{d \Tilde{r}^2} + \left[\left(1- \frac{2}{\tilde{l}_B^2}\right) \Tilde{r} - 2 \left(m+\frac{1}{2}\right)\right] \frac{d p^\uparrow_{\perp}}{d \Tilde{r}}\\
&&+ E^2 p_{\perp}^\uparrow = 0 \,, \notag
\end{eqnarray}
\begin{eqnarray} \label{Eq:AnsatzForHigherLLPolynomialCoax}
&&\Tilde{r}^2 \frac{d^2 p^\uparrow_{\parallel}}{d \Tilde{r}^2} + \left[-\frac{1}{\Tilde{l}_B^2}\Tilde{r}^2 + \Tilde{r} - 2 \left(m + \frac{1}{2} \right)\right] \frac{d p^\uparrow_{\parallel}}{d \Tilde{r}}\\
&&+ E^2 p_{\parallel}^\uparrow = 0 \, .  \notag
\end{eqnarray}
For non-negative $m$ in case of the coaxial field, a similar ansatz for the down component with reversed $m$, turns out to yield meaningful results:
\begin{equation}\label{Eq:AnsatzForHigherLLcoaxMpos}
\Psi_{\parallel}^{\downarrow}(\Tilde{r}) = p_{\parallel}^{\downarrow}(\Tilde{r}) \, \Tilde{r}^{-\frac{1}{2}} \, e^{-\frac{\tilde{r}}{2 \Tilde{l}_B^2}} \, e^{-\frac{1}{\Tilde{r}} \left(m+\frac{1}{2}\right)} \,.
\end{equation}
Inserting this into Eq.~\eqref{EQ_-} yields
\begin{eqnarray} \label{Eq:AnsatzForHigherLLPolynomialCoaxMpos}
&&\Tilde{r}^2 \frac{d^2 p^\downarrow_{\parallel}}{d \Tilde{r}^2} + \left[-\frac{1}{\Tilde{l}_B^2}\Tilde{r}^2 + \Tilde{r} + 2 \left(m + \frac{1}{2} \right)\right] \frac{d p^\downarrow_{\parallel}}{d \Tilde{r}}\\
&&+ \left(E^2 - \frac{2 \left(m + \frac{1}{2} \right)}{\Tilde{l}_B^2} - \frac{1}{\Tilde{l}_B^2} \Tilde{r} \right) p_{\parallel}^\downarrow = 0 \, . \notag
\end{eqnarray}

\begin{figure}[tbp]
  \centering
  \includegraphics[width=0.8\columnwidth]{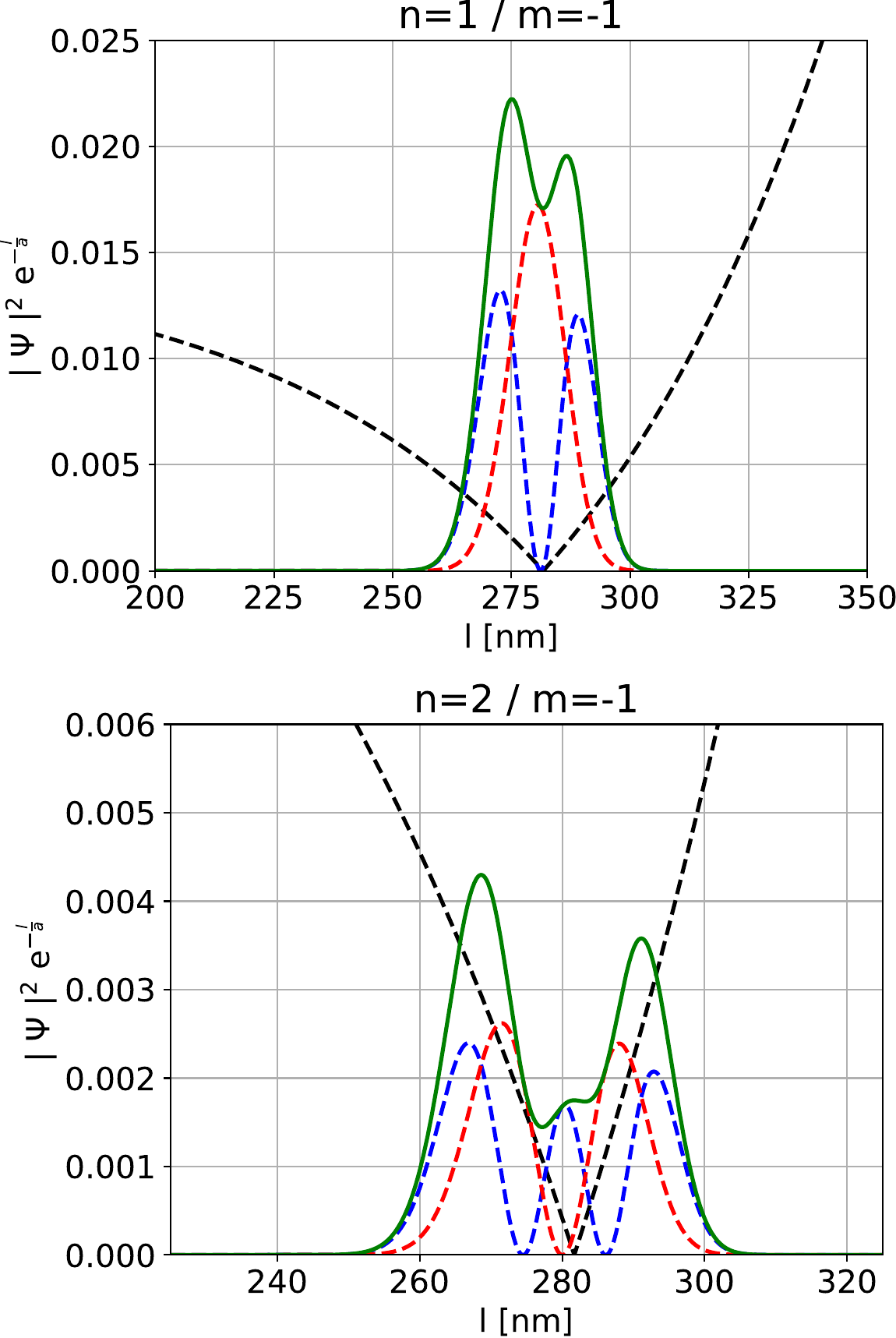}

  \caption{Pseudosphere radial probability densities \(\abs{\Psi}^2 e^{-\frac{l}{a}}\) for the up component (blue, dashed lines), down component (red, dashed lines) and the whole spinor (green, solid  lines) for the first and second excited states of $\hat{H}_{\perp}$, Eq.~(\ref{Eq:DiracForMinding}), with $m=-1$.
  Maxima of the wavefunction probabilities (not-normalized) are localized near the minima of the corresponding 
  effective potentials 
  \(\abs{V_\perp(l)}\) (black, dashed lines).
  The radius of curvature \(a=60\)\,nm and the
  strength of the perpendicular magnetic field \(B=10\)\,T.}
  \label{FIG_WF12}
\end{figure}



\subsubsection{Higher Landau levels: perpendicular magnetic field}

The solutions of Eq.~(\ref{Eq:AnsatzForHigherLLPolynomial}) can be expressed in terms of the \emph{associated Laguerre polynomials} of order $n \geq 0$ as shown in Appendix~\ref{APP_RADIAL}. This constrains the square of the (dimensionless) energy $E$ to satisfy
\begin{equation}
E_n^2 = n\,\left(\frac{2}{\tilde{l}_B^2} - n \right) = 
n\,\left(\frac{\Phi_\perp}{\Phi_0} - n \right) \,.
\end{equation}
Since only real eigenenergies are physical, we get an upper limit for the principal quantum number $n$ defining the order of Laguerre polynomial. This limit is given by 
$\lfloor 2a^2/l_B^2 \rfloor = \lfloor \Phi_\perp/\Phi_0 \rfloor$, {\it i.e.}~by a \textit{ratio of the squares of two characteristic length scales---the curvature radius $a$ and the magnetic length $l_B$.}

In view of the chiral symmetry, Eq.~(\ref{EQ:symmetry1}), the spectrum is symmetric with respect to zero, and hence the scaled eigenenergies \(E_n\) including the zero-mode read
\begin{equation}
    E_n = \pm \sqrt{n \left(\frac{\Phi_\perp}{\Phi_0} - n \right)}, \,\ \ n \in \left\{0, \dots, \left\lfloor \frac{\Phi_\perp}{\Phi_0} \right\rfloor\ \right\}\,.
    \label{eq:n-limit}
\end{equation}

The up components of the corresponding eigenfunctions read
\begin{equation}\label{EQ:UpComp}
\Psi^\uparrow_{\perp, n}(\Tilde{r}) = \Tilde{r}^n L_n^{(-2n+2/\Tilde{l}_B^2)} \left(-\frac{2m + 1}{\Tilde{r}} \right) \Psi^\uparrow_{\perp, 0}(\Tilde{r}) \, .
\end{equation}

The down components are calculated by  explicitly evaluating Eq.~\eqref{Eq:FirstAndSecond2},
\begin{equation}
    \Psi^\downarrow_{\perp, n}(\Tilde{r}) = \frac{1}{E_n}\, \hat{L}^+_{\perp}\, \Psi^\uparrow_{\perp, n}(\Tilde{r}) \, ,
\end{equation}
yielding again solutions in terms of an associated Laguerre polynomial:
\begin{eqnarray}\label{EQ:DownComp}
\Psi^\downarrow_{\perp, n}(\Tilde{r}) &=& i \frac{1}{E_n} \left(-n+\frac{2}{\Tilde{l}_B^2}\right) \Tilde{r}^{n} \\
&& \times L_{n-1}^{(-2n+2/\Tilde{l}_B^2)} \left(-\frac{2m+1}{\Tilde{r}}  \right) \Psi^\uparrow_{\perp, 0}(\Tilde{r}) \, . \notag
\end{eqnarray}
Solutions for negative magnetic fields follow from the $\sigma_y \mathscr{C}$ symmetry,
Eq.~(\ref{EQ:symmetry2}). The latter exchanges up and down components of the wavefunctions 
and reverses the signs of the angular momentum quantum number $m$; otherwise, the spectrum remains the same.

To summarize, independent of the sign of the magnetic field, the eigenenergies 
are degenerate in $m$ and are given by (in units of 
$\mathcal{E}=\hbar v_F/a$)
\begin{equation}\label{EQ_DISP}
    E_n = \pm \sqrt{n \left(\frac{\abs{\Phi_\perp}}{\Phi_0} - n \right)} \, , \, \ \ n \in \left\{0,\dots, \left\lfloor \frac{\abs{\Phi_\perp}}{\Phi_0} \right\rfloor\ \right\} \, .
\end{equation}
However, as mentioned above, physically allowed solutions only exist for $m \in \mathcal{M}_{\perp}$.

A figure of this spectrum is shown in the numerics section of this paper.

Correspondingly, Fig.~\ref{FIG_WF12} displays the radial probability densities for the wavefunctions of the first two excited states with \(m=-1\). It can be seen that for positive $B$ the up component of the \(n\)-th excited state has \(n\) zeros---associate Laguerre polynomial of the $n$-th degree---and the down 
component \((n-1)\) zeros---associate Laguerre polynomial of the $(n-1)$-th degree. Furthermore, it seems that the wavefunctions become spatially more broadened upon rising the principal quantum number \(n\). We take this observation as numerical evidence without any rigorous proofs for a generic \(n\).

\subsubsection{Higher Landau levels: coaxial magnetic field}

The experimentally more relevant case of a homogeneous coaxial magnetic field is more complicated since the $B$-field appears inhomogeneous on the surface of the pseudosphere and its perpendicular component is arclength-dependent.
It turns out that the only relevant parameter in this case is
\begin{eqnarray}
\eta &=&  \frac{|2m+1|}{l_B^2} = |2m+1| \frac{|\Phi_{\parallel}|}{\Phi_0}.
\end{eqnarray}
We compute the higher Landau levels in two ways.

First, as shown in Appendix~\ref{APP_RADIAL} the eigenfunctions
of Eq.~(\ref{Eq:AnsatzForHigherLLPolynomialCoax}) and Eq.~(\ref{Eq:AnsatzForHigherLLPolynomialCoaxMpos}) can be expressed by means of the \emph{double confluent Heun functions}. Since it is a bit cumbersome to work with these functions, we do not give explicit expressions for the underlying wavefunctions here; rather, we discuss the spectrum. 
Using asymptotic methods developed for the double confluent Heun equation on the complex-plane, see \cite{Slavianov1996-fr, W_Lay_1998} and references therein, the squares of eigenenergies $E_n$ for $\tilde{l}_B^2>0$ can be estimated by successive approximations, to the second order in the semiclassical parameter $1/\eta$. For negative $m$, see Eq.~(\ref{EQ_+}), they are given by the following expression (normalized to $\mathcal{E}=\hbar v_F/a$):
\begin{equation}
E_{n,m}^2 \approx 2n \sqrt{\eta} + \frac{1}{2}n^2 -1 \, .
\end{equation}
For non-negative $m$, see Eq.~(\ref{EQ_-}), they are given by
\begin{equation}
E_{n,m}^2 - \eta \approx \left(2n + 1\right) \sqrt{\eta} + \frac{1}{2}n^2 + \frac{1}{2} n -2 \, .
\end{equation}
The approximation works for high values of the argument under the square root, i.e., large $|m|$ and large magnitude of $B$.
In accordance with the chiral symmetry, Eq.~(\ref{EQ:symmetry1}), the eigenenergies are distributed symmetrically around zero energy.
\begin{eqnarray} \label{EQ:SpectrumPseudoSphereCoaxial}
E_{n,m} &=& \pm \sqrt{2n \sqrt{\eta} +\frac{1}{2} n^2 -1} \\
& \underset{n \gg 1}{\approx}  & 
\pm \sqrt{\frac{n}{2} \left(4\sqrt{\eta}+n\right)} \, , \notag 
\end{eqnarray}
and for non-negative $m$

\begin{eqnarray} \label{EQ:SpectrumPseudoSphereCoaxialMpos}
E_{n,m} &=& \pm \sqrt{\eta + (2n+1) \sqrt{\eta} +\frac{1}{2} n^2 + \frac{1}{2} n -2} \\
& \underset{n \gg 1}{\approx}  & 
\pm \sqrt{\eta + (2n+1) \sqrt{\eta} +\frac{1}{2} n^2 + \frac{1}{2} n} \, . \notag 
\end{eqnarray}

Here, it is worth to spell out several observations: 

(i) As can be seen, the approximate formula for $E_{n,m}$ does not include the zero energy eigenstates that are degenerate in $m$. 

(ii) Contrary to that and also to the case of an overall perpendicular field the higher Landau levels for $\vec{B}_\parallel$ are no longer degenerate in $m$. 
This can be related to the fact that the eigenfunctions for different $m$ are spatially centered near different $l$ -- the minima of $|V_\parallel|$ -- and the field strength component of $\vec{B}_\parallel$ perpendicular to the surface $\Sigma$ at such $l$, \textit{i.e.}~$(\vec{n}\cdot\vec{B}_\parallel)$, varies substantially with the arclength coordinate~\footnote{Lifting the Landau level degeneracy is also possible in presence of a non-uniform electrical field \cite{Edery2019}.}.

(iii) Most notably, for strong (positive) magnetic fields, $E_{n,m} \propto |B|^{1/4}$ for $m < 0$ and $E_{n,m} \propto |B|^{1/2}$ for $m \geq 0$. This peculiar parametric $B$-dependence should be experimentally observable. For one half of the spectrum, it characteristically differs from the usual linear scaling with $|B|$ in the planar Schrödinger case, and also from the characteristic $\sqrt{|B|}$-dependence for the corresponding Dirac case in constant perpendicular field.

(iv) Moreover, due to the ``$+$''-sign in front of the term $n^2/2$, and contrary to the case of $\vec{B}_\perp$, Eq.~(\ref{eq:n-limit}), $n$ has no formal upper 
limit. However, when demanding that the states are localized within the surface, it becomes clear that the eigenenergies must not be larger than the confining effective potential, which sets a natural upper bound for $n$.

Second, we obtain the spectrum making a WKB analysis, see App.~\ref{SEC_WKB}. We find that in the limit of strong magnetic fields, the spectrum is approximately given by (see also Eq.~(\ref{DISP_WKB}))
\begin{eqnarray}
E_{n,m}  &\approx& \sqrt{\eta + \text{sgn} \left(m + \frac{1}{2} \right) \eta + \pi \left(n + \frac{1}{2} \right) \sqrt{\eta}} \, . \notag 
\end{eqnarray}

Again, a few remarks are worth to be mentioned:

(i) As in the formula obtained from Heun asymptotics, the zero energy states are not covered.

(ii) For $m < 0$, the first two terms under the square root cancel and the asymptotic behaviour for strong magnetic fields is again given by $E_{n, m} \propto |B|^{1/4}$.

(iii) For $m \geq 0$, the first two terms add and are dominant in the strong field limit, yielding $E_{n, m} \propto |B|^{1/2}$.

(iv) The upper limit for the energy relies on the assumption made during the WKB treatment that the two classical turning points are at a radius smaller than $a$.

\subsection{Minding surface}
\label{sec:Minding}

In the following we sumarize the main spectral findings for the Minding surface depicted in Fig.~\ref{FIG_KAPPA}. 

\subsubsection{Zeroth Landau level }

Considering $\gamma\neq 0$ in the Eqs.~(\ref{Eq:FirstAndSecond1})~and~(\ref{Eq:FirstAndSecond2}) for the effective potentials $V_{\perp/\parallel}$ and the auxiliary operators $L^{\pm}_{\perp/\parallel}$ leads to slightly more involved radial  
Eq.~(\ref{eq:Lpm}).
Introducing the auxiliary variable $\rho=\arcsin{(\Tilde{r}/\Tilde{\gamma})}$ the expressions get simplified, and after some math the full spinorial solutions
$\bigl( \Psi^\uparrow(\Tilde{r},\phi),\Psi^\downarrow(\Tilde{r},\phi) \bigr)^\top$ for the zeroth Landau level on the Minding surface for perpendicular and coaxial fields read
\begin{align}
\Psi^+_{0, \perp} &= \Tilde{r}^{-\frac{1}{2} - \frac{1}{\tilde{l}_B^2}} e^{+\frac{1}{\Tilde{\gamma}} \left(m+\frac{1}{2}\right) \arcsin \left(\frac{\Tilde{\gamma}}{\Tilde{r}}\right)+i \left(m+\tfrac{1}{2} \right)\phi}
\begin{pmatrix}
1 \\
0
\end{pmatrix}\,,\\
\Psi^-_{0, \perp} &= \Tilde{r}^{-\frac{1}{2} + \frac{1}{\tilde{l}_B^2}} e^{-\frac{1}{\Tilde{\gamma}} \left(m+\frac{1}{2}\right) \arcsin \left(\frac{\Tilde{\gamma}}{\Tilde{r}}\right)+i \left(m+\tfrac{1}{2} \right)\phi}
\begin{pmatrix}
0 \\
1
\end{pmatrix}\,.
\end{align}
\begin{align}
\Psi^+_{0, ||} &= \Tilde{r}^{-\frac{1}{2}} e^{-\frac{\sqrt{\Tilde{r}^2 - \Tilde{\gamma}^2}}{2\Tilde{l}_B^2}} e^{+\frac{1}{\Tilde{\gamma}} \left(m+\frac{1}{2}\right) \arcsin \left(\frac{\Tilde{\gamma}}{\Tilde{r}}\right)+i \left(m+\tfrac{1}{2} \right)\phi}
\begin{pmatrix}
1 \\
0
\end{pmatrix}\,,\\
\Psi^-_{0, ||} &= \Tilde{r}^{-\frac{1}{2}} e^{+\frac{\sqrt{\Tilde{r}^2 - \Tilde{\gamma}^2}}{2\Tilde{l}_B^2}} e^{-\frac{1}{\Tilde{\gamma}} \left(m+\frac{1}{2}\right) \arcsin \left(\frac{\Tilde{\gamma}}{\Tilde{r}}\right)+i \left(m+\tfrac{1}{2} \right)\phi}
\begin{pmatrix}
0 \\
1
\end{pmatrix}\,.
\end{align}
Since Minding surfaces do not behave singularly and $r_{min}=\gamma$, both solutions are square-integrable. Moreover, in the limit
$\gamma \to 0$ one indeed recovers the zeroth Landau level solutions for the pseudosphere given by 
Eqs.~(\ref{Eq:PseudoSpehereZeroLandauLevelUp})~,~(\ref{Eq:PseudoSpehereZeroLandauLevelDown})~,~(\ref{Eq:PseudoSpehereZeroLandauLevelUpCoax})~and~(\ref{Eq:PseudoSpehereZeroLandauLevelDownCoax}).


\subsubsection{Higher Landau levels}

Unfortunately, employing the same techniques for the Minding surface as for the pseudosphere does not yield any closed-form solution. Therefore, we resort to numerics. As shown and discussed below,
we find strong numerical evidence that the energies of higher Landau levels for the Minding surface in $\vec{B}_\perp$ are given by the same expression as Eq.~(\ref{EQ_DISP}) for the pseudosphere.

This is in accordance with the fact that the area of the embeddable part of the Minding surface with radius of curvature $a$ does not depend on $\gamma$ and becomes equal to the area of the pseudosphere possessing the same radius $a$ and, hence, the total enclosed flux $|\Phi|=2\pi a^2 |B|$.


\section{Numerical calculations}\label{Numerical}

Our analytical results for the pseudosphere in a coaxial field are only asymptotically exact, and we did not manage to solve the Dirac equation for the Minding surface in 
$\vec{B}_{\perp/\parallel}$ for non-zero energy states.  We thus perform numerical calculations
following the procedure laid out in Refs.~\cite{Kozlovsky2020,GRAF}, adapted to the present context.
For the sake of a self-contained presentation we recall its essentials.

\subsection{Method and implementation}

The continuous Dirac equation, Eq.~(\ref{EQ_DIRAC_I}), is discretized on a rectangular lattice and treated by means of the tight-binding package \emph{Kwant} \cite{E}. The lattice is thus purely numerical, i.e. it does not represent the underlying atomic lattice of the TI sample – recall the discussion in the introduction.
To deal with operators and fields in curvilinear coordinates, the tight-binding-generating  hoppings are non-homogeneous and depend on entries of the metric tensor $g_{\mu\nu}(\phi,r)=\text{diag}(r^2,\frac{a^2}{r^2-\gamma^2})$. One way to simplify this is to use the 
angle-arclength coordinates $(\phi, l)$. Transforming from $(\phi,r)$ to $(\phi,l)$ simplifies the spatial-metric tensor from $g_{\mu\nu}(\phi,r)=\text{diag}(r^2,\frac{a^2}{r^2-\gamma^2})$ 
to $g_{\mu\nu}(\phi,l)=\text{diag}(r^2(l),1)$.
Consequently, hoppings along the $l$ direction on the underlying lattice do not depend on the current surface position, in contrast to the $(\phi, r)$ parameterization. Of course for both parameterizations, the hoppings along the $\phi$ coordinate are position-dependent.

Changing $r$ to $l$, according to Eq.~(\ref{Eq:rVsl}), transforms the Hamiltonians $\hat{H}_{\perp/\parallel}$, 
Eqs.~(\ref{Eq:DiracForMinding})~and~(\ref{Eq:DiracForMindingCoax}), and also the 
underlying volume form 
\begin{equation}
\omega_g=\sqrt{|\text{det}\,g_{\text{space}}|}\,d\phi\wedge d l= r(l)\, d\phi\wedge d l \, .
\end{equation}
Denoting the transformed Hamiltonian as $\hat{H}(\phi,l)$, 
the differential operators with respect to $l$ and $\phi$ entering it should be substituted 
by corresponding operators for finite differences on a discrete lattice.
Using the symmetric difference scheme inherent to the flat Euclidean space leads to hermitian matrices in the conventional sense~\cite{A}. However, applying that prescription to $\hat{H}(\phi,l)$ would lead to a problem -- the matrix of the discretized Hamiltonian $\hat{H}(\phi,l)$ would not be hermitian. 
Hence, the hermiticity with respect to the curved volume form $\omega_g = r(l)\, d\phi\wedge d l$ is required:
\begin{align}
    \langle\psi_1,\hat{H}\psi_2\rangle &=\int dl d\phi\ r(l)\ (\overline{\psi}_1)^\mathcal{I}\ (\hat{H}\psi_2)_\mathcal{I} \\
    &= \int dl d\phi\ r(l)\ \overline{(\hat{H}\psi_1)^\mathcal{I}}\ (\psi_2)_\mathcal{I} \\
     &=\langle\hat{H}\psi_1,\psi_2\rangle\,.\notag
\end{align}
We circumvent this issue by redefining the spinor fields and, accordingly, the Hamiltonian $\hat{H}(\phi,l)$ such that
\begin{align}
    &\langle\psi_1,\hat{H}\psi_2\rangle =\int dl d\phi\ r(l)\ (\overline{\psi}_1)^\mathcal{I}\ (\hat{H}\psi_2)_\mathcal{I} \notag\\
    &= \int dl d\phi\ \overline{(\sqrt{r(l)}\psi_1)^\mathcal{I}}\ \ \Bigl(\sqrt{r(l)} \hat{H} \frac{1}{\sqrt{r(l)}}\ \sqrt{r(l)} \psi_2\Bigr)_\mathcal{I} \notag\\
    &=\int dl d\phi\ (\overline{\widetilde{\psi}}_1)^\mathcal{I}\ (\widetilde{H}\widetilde{\psi}_2)_\mathcal{I}\,.
\end{align}
The appropriately rescaled Hamiltonian
\begin{align}\label{Eq:RescaledDiracHam}
    \widetilde{H}(\phi,l)=\sqrt{r(l)}\,\hat{H}(\phi,l)\,\frac{1}{\sqrt{r(l)}}
\end{align}
can now be routinely discretized by the standard symmetric difference scheme leading to a hermitian matrix in the conventional (flat space) fashion.
In what follows we calculate the spectra of such rescaled Hamiltonians for the pseudosphere and Minding surfaces. 
We also add a Wilson mass term to the Hamiltonian $\widetilde{H}$ in order to avoid fermion doubling,
for details see \cite{A}. Furthermore we use the hard-wall boundary conditions along the $l$ direction (see discussion below) and anti-periodic boundary conditions along $\phi$. For the sake of compactness 
we do not provide the explicit expressions of the rescaled Hamiltonians.

The embeddable part of the pseudosphere is non-compact. Thus we can apply the discretization only to a certain finite domain $\mathscr{D}$ of $\Sigma$. 
For the latter we choose a part close to the rim (trumpet) of the pseudosphere, i.e.~$\mathscr{D}$ covers the part of radii descending from $r_{max}=a$ up to some end radius $r_e$, or in the arclength parameterization, the lengths from $l=0$ to $l_e=a\ln{(a/r_e)}$.
So the discrete space is a 2D rectangular grid within \([0, 2 \pi] \times [0, l_e]\). 
We use \(N_W\) lattice points sampling the interval $[0,2\pi]$ and fix the number of lattice points 
in $[0,l_e]$ by introducing a lattice constant \(b\), see Fig.~\ref{FIG_WF_NUM}.
Using the symmetric difference approximation 
the discretization of the derivatives along the $\phi$ and $l$ directions is given by
\begin{eqnarray}
\frac{d\tilde{\Psi}_{i, j}}{d\phi} \approx \frac{\tilde{\Psi}_{i+1, j}-\tilde{\Psi}_{i-1, j}}{2 \frac{2\pi}{N_W}}, \\
\frac{d\tilde{\Psi}_{i, j}}{dl} \approx \frac{\tilde{\Psi}_{i, j+1}-\tilde{\Psi}_{i, j-1}}{2b},
\end{eqnarray}
where $i$ and $j$ numerate the lattice points in $\phi$ and $l$ direction, respectively.

In order to compare analytical and numerical results we consider a
pseudosphere and a Minding surface with the radius of curvature $a=60$\,nm, and, correspondingly, with $\gamma=0$ and $\gamma = 1$\,nm \footnote{The results can be simply rescaled (in length and energy) to obtain energies, $B$-fields strengths and length scales for other sizes}.


\subsection{Pseudosphere:
numerical vs. analytical results}

For the numerics we use, besides $a=60$\,nm, as minimum radius $r_e=0.25$\,nm, corresponding to $l_{e}=328.8$\,nm. 
This assumption is justified for wavefunctions decaying sufficiently fast towards the conically narrowing pseudosphere singularity at
$r \to 0$ ($ l \to \infty$), as anticipated by our analytical ansatz, see Eq.~(\ref{Eq:AnsatzForHigherLL}). 

Employing hard-wall boundary conditions at $l=l_{e}$ and also at $l=0$ 
we expect certain deviations between the numerical and analytical spectra for small magnetic fields, as can be anticipated from the shapes and positions of the effective potentials $|V_{\perp/\parallel}(l)|$ shown in Figs.~\ref{FIG_VPOT}~and~\ref{FIG_VPOT_COAX}. For weak fields the potential minima approach the rim of the pseudosphere and, correspondingly, the hard-wall boundary conditions at $l=0$ will affect the numerically-computed eigendata, as opposed to analytical solutions.
However, for sufficiently large fields, when the minima of $|V_{\perp/\parallel}|$ are pronounced and sufficiently far away from the rim, the numerical and analytical results should coincide and the finite-size effects stemming from the boundary-conditions would diminish. 

\begin{figure}[tbp]
  \centering
  \includegraphics[width=0.8\columnwidth]{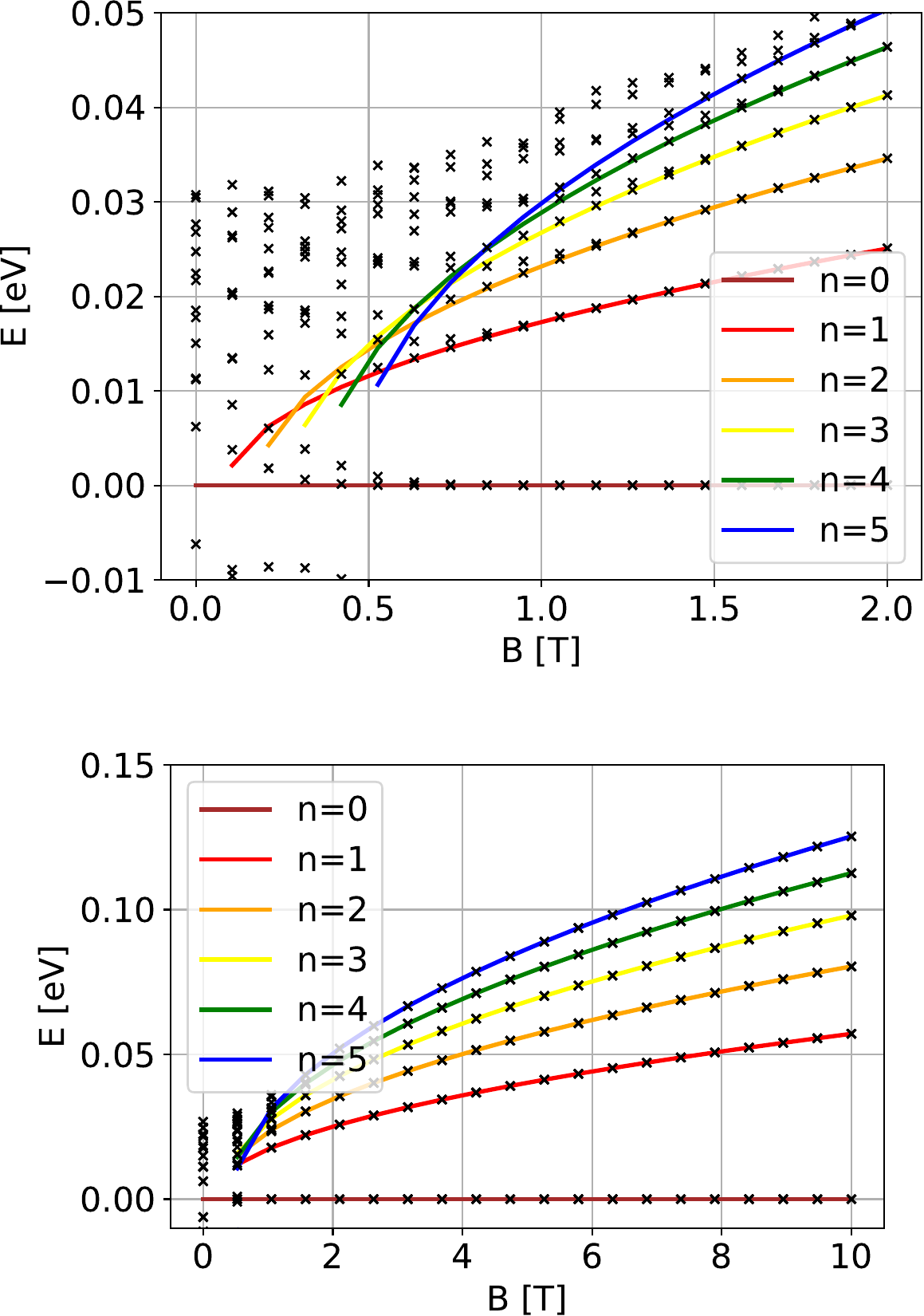}
  
  \caption{Landau level spectrum of the pseudosphere in an overall perpendicular field $\vec{B}_\perp$ of strength $B$.
  Low-lying eigenenergies $E_n(B)$ obtained analytically, Eq.~(\ref{EQ_DISP}), solid lines
  (color indicates Landau level quantum number $n$) are compared to numerical data (black symbols).
  Panel (a) shows a zoom into the low-field regime of the spectrum shown in panel (b) for larger $B$-scales.
  Numerical parameters used: radius of curvature 
  \(a=60\)\,nm, number of angular grid points \(N_W=8\), and lattice spacing \(b=0.5\)\,nm.
  Note that not all allowed higher-lying eigenvalues are depicted  and, for the sake of clarity, only zero and positive eigenvalues are displayed.
  }
  \label{FIG_DISP_NUM}
\end{figure}

We start with the case of the perpendicular magnetic field $\vec{B}_\perp$. 
In Fig.~\ref{FIG_DISP_NUM} we compare the numerically computed spectrum (as a function of field strength $B$) with the analytical results based on Eq.~(\ref{EQ_DISP}) when varying the magnitude $B$ of $\vec{B}_\perp$.
For the chosen parameter range we find that for $B > 2$\,T the numerical eigenenergies (black symbols) agree very well with the corresponding analytical results (solid lines).
As expected, upon
increasing the magnetic field, the numerical eigenvalues rapidly converge to the analytical Landau levels $E_n$ given by Eq.~(\ref{EQ_DISP}), in the order given by their principal quantum number $n$. 
For $B < 2 T$ we see substantial deviations in the energy spectrum attributed to the effects of boundary-conditions.

Furthermore, Fig.~\ref{FIG_WF_NUM} shows numerically computed radial probabilities for three 
eigenstates $\bigl( \Psi^\uparrow,\Psi^\downarrow \bigr)^\top$ of the discretized Dirac Hamiltonian $\widetilde{H}$ along with their analytical counterparts, given by Eqs.~(\ref{EQ:UpComp})~and~(\ref{EQ:DownComp}). 
Also these results corroborate the excellent matching between numerics and analytics at 
high(er) magnetic fields.\\


\begin{figure}[tbp]
  \centering
  \includegraphics[width=1\columnwidth]{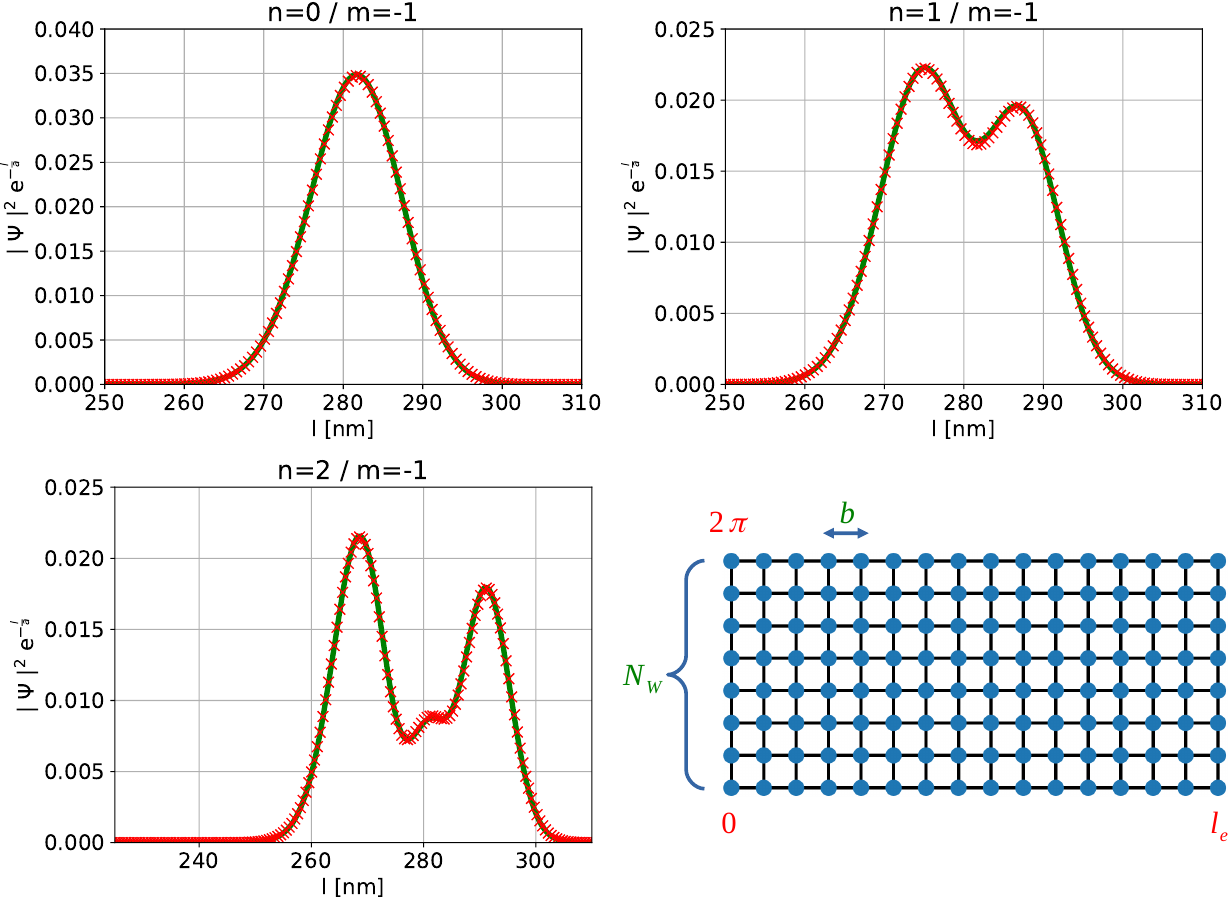}

  \caption{Radial probability densities (unnormalized) for a pseudosphere in perpendicular magnetic field $\vec{B}_\perp$. 
  Analytically obtained results, \(\abs{\Psi}^2 e^{-\frac{l}{a}}\), (solid, green lines) are compared to the corresponding numerical solutions (red crosses) for the three lowest Dirac Landau levels of
  $\hat{H}_\perp$ with the angular quantum number \(m=-1\).
  Numerical parameters: \(N_W=20\), \(b=0.5\)\,nm, pseudosphere radius 
  \(a=60\)\,nm, the field strength \(B=10\)\,T. \textendash{} Schematic of the lattice used in the discretization of 
  $\hat{H}_{\perp/\parallel}$ with \(N_W \) points along the angular direction, and length $l_e$ and lattice constant \(b\) along the horizontal arclength direction.}
  \label{FIG_WF_NUM}
\end{figure}

For the coaxial field $\vec{B}_\parallel$ we employ the same discretization method. Here, however, the eigenenergies are not degenerate in the angular quantum number $m$, so fitting the numerical data with the help of the asymptotic eigenenergy formula, 
Eq.~(\ref{EQ:SpectrumPseudoSphereCoaxial}), would be quite messy and non-illustrative due to a substantial number of scattered points.
For this reason we decided to only implement the radial equations for 
$\bigl( \Psi^\uparrow,\Psi^\downarrow \bigr)^\top$, see Eqs.~(\ref{Eq:FirstAndSecond3})~and~(\ref{Eq:FirstAndSecond4}), for which the angular number $m$ enters via the effective 
wedge-potential $|V_\parallel|$, consult Eq.~(\ref{eq:Vcoax}) and Fig.~\ref{FIG_VPOT_COAX}. 

First of all we compare the numerical results to the spectrum from Heun asymptotics, Eq.~(\ref{EQ:SpectrumPseudoSphereCoaxial}). Figure~\ref{FIG_DISP_NUM_COAX} shows the resulting numerical spectra for the quantum numbers $m=-1$ and $m=0$ as functions of the strength $B=||\vec{B}_\parallel||$. The matching with the numerically obtained spectra becomes very good and improves, correspondingly, for larger magnetic fields and higher angular number $|m|$.


\begin{figure}[tbp]
  \centering
  \includegraphics[width=0.8\columnwidth]{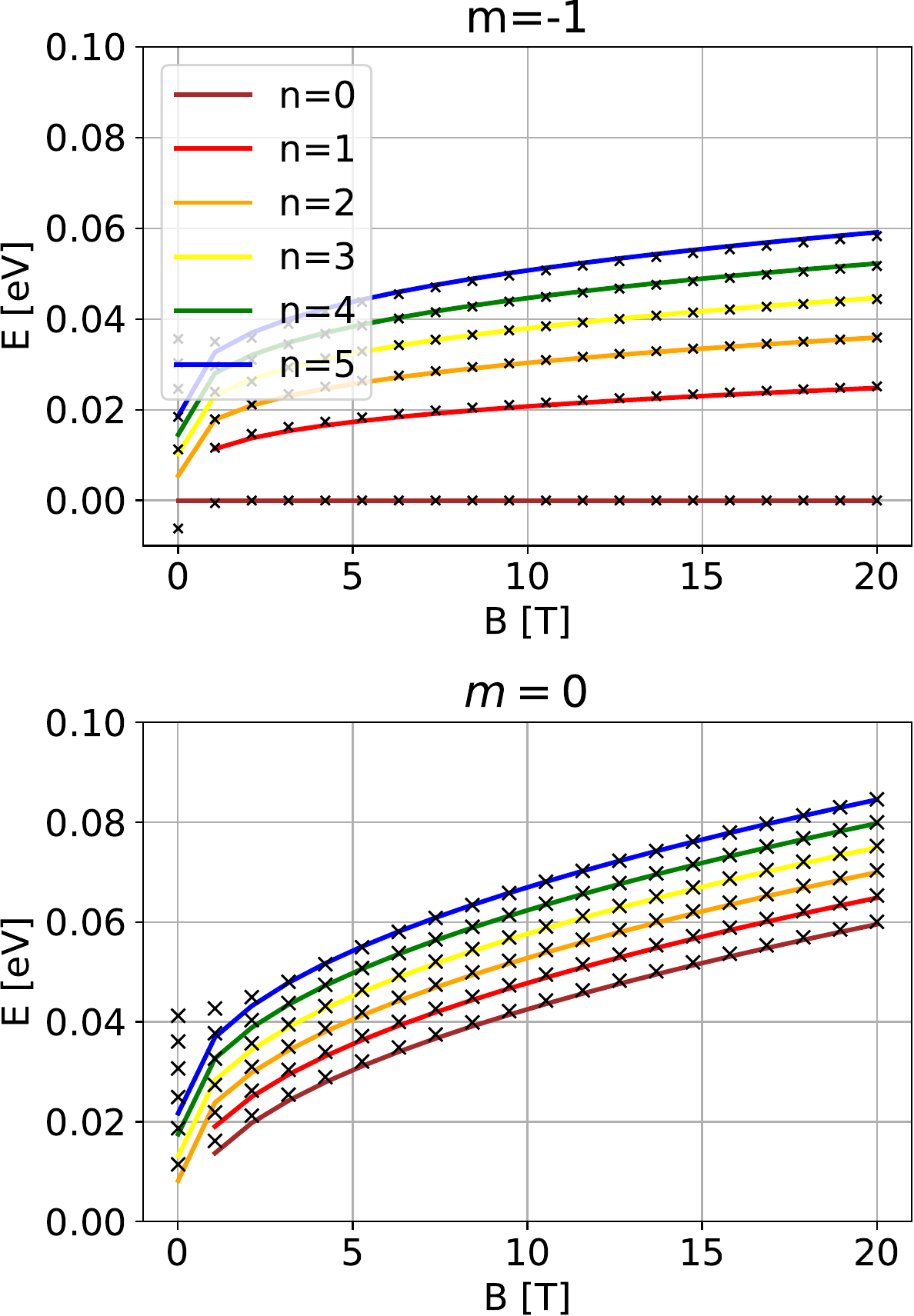}
  \caption{
  Landau level spectrum of the pseudosphere in a coaxial field $\vec{B}_\parallel=(0,0,B)$.
  Lowest eigenenergies as functions of the magnetic field $B$ obtained analytically, 
  Eq.~(\ref{EQ:SpectrumPseudoSphereCoaxial}) and Eq.~(\ref{EQ:SpectrumPseudoSphereCoaxialMpos}), (solid, colored lines labeled by principal quantum number) are compared with numerical solutions (black symbols) of Eq.~(\ref{Eq:FirstAndSecond3}) for $m=-1$ (a) and $m = 0$ (b).
  Here: the pseudosphere radius \(a=60\)\,nm, and the arclength lattice constant \(b=0.5\)\,nm, 
  for the sake of brevity, only zero and positive eigenvalues are displayed.}
  \label{FIG_DISP_NUM_COAX}
\end{figure}

We also compare with the WKB approximation. As can be observed from Fig.~\ref{FIG_DISP_NUM_COAX_WKB}, especially for negative $m$, the levels obtained from WKB seem 
to be a bit shifted compared to the numerical results. However, the asymptotical behaviour seems to be correct. Contrary to this, for non-negative $m$, the matching is very good.

\begin{figure}[tbp]
  \centering
  \includegraphics[width=0.8\columnwidth]{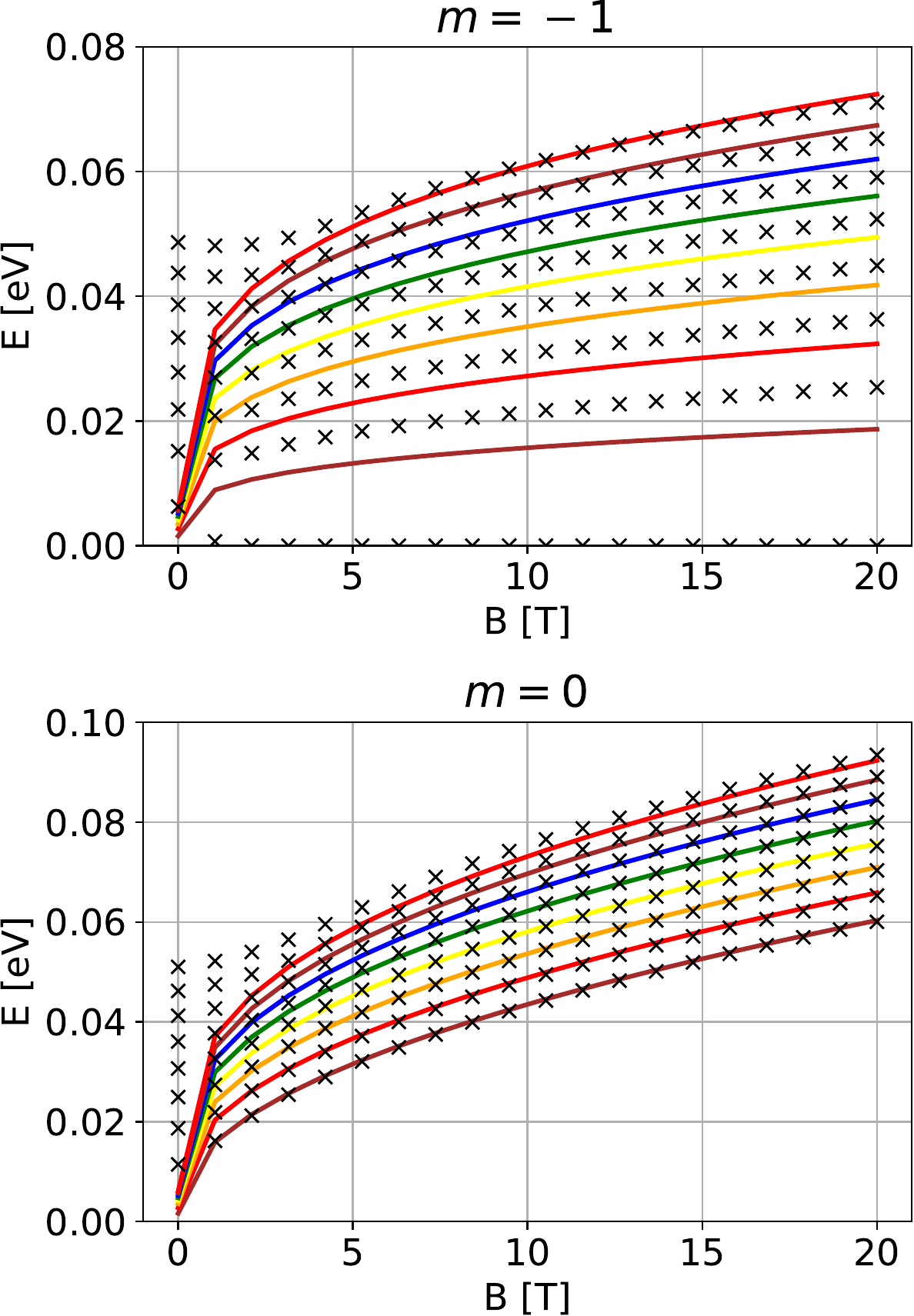}
  
  \caption{
  Landau level spectrum of the pseudosphere in a coaxial field $\vec{B}_\parallel=(0,0,B)$.
  Lowest eigenenergies as functions of the magnetic field $B$ obtained by means of WKB methods, Eq.~(\ref{DISP_WKB}), (solid, colored lines labeled by principal quantum number) are compared with numerical (black symbols) solutions of Eq.~(\ref{Eq:FirstAndSecond3}). 
  Panel~(a) corresponds to states with $m=-1$ displaying $|B|^{1/4}$ asymptotic, while panel~(b) represents states with $m=0$ with $|B|^{1/2}$ asymptotic.
  Here: the pseudosphere radius \(a=60\)\,nm, and the arclength lattice constant \(b=0.5\)\,nm, 
  for the sake of brevity, only zero and positive eigenvalues are displayed.
  }
  \label{FIG_DISP_NUM_COAX_WKB}
\end{figure}

Figure~\ref{FIG_WF_NUM_COAX} shows the corresponding wavefunctions for a few Landau levels in the coaxial field. As explained before, we only provide numerical solutions for the associated eigenfunctions, the analytical treatment being fairly complicated. Nevertheless, the probability maxima of the eigenstates are developing close to positions, where the effective wedge-potentials $|V_\parallel|$ become minimized, see Fig.~\ref{FIG_VPOT_COAX}. It can also clearly be seen that states with negative angular quantum number (e.g. $m=-1$) differ from states with non-negative angular quantum number (e.g. $m=0$).

\begin{figure}[tbp]
  \centering
  \includegraphics[width=1\columnwidth]{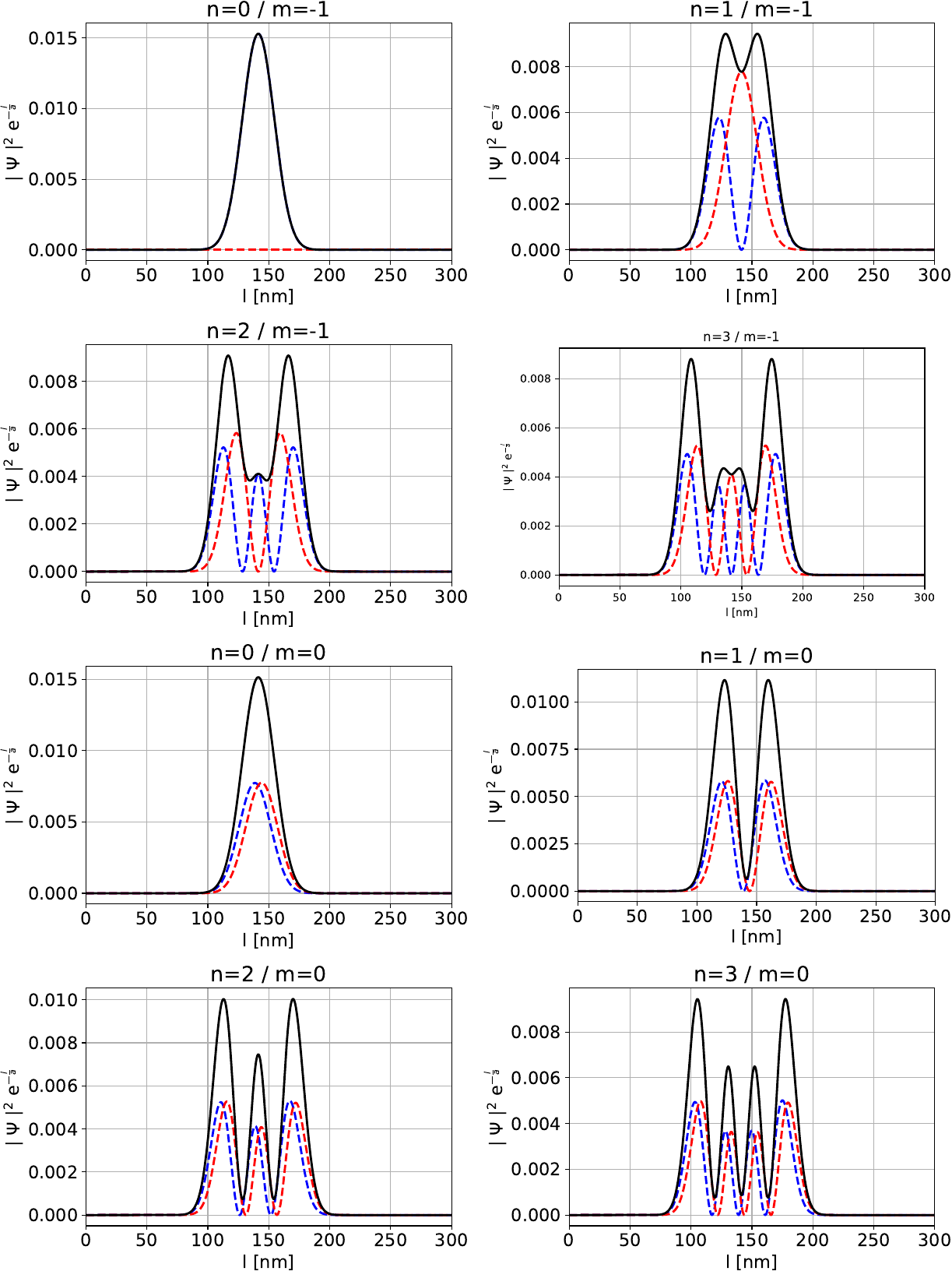}

  \caption{Numerically gained pseudosphere radial probability densities \(\abs{\Psi}^2 e^{-\frac{l}{a}}\) for the up component (blue, dashed lines), down component (red, dashed lines) and the whole spinor (black, solid  lines) for the lowest four states of $\hat{H}_{||}$, with $m=-1$ and $m=0$.
  Maxima of the wavefunction probabilities (not-normalized) are localized near the minima of the corresponding 
  effective potentials 
  \(\abs{V_{||}(l)}\).
  The radius of curvature is \(a=60\)\,nm and the
  strength of the perpendicular magnetic field equals \(B=20\)\,T.}
  \label{FIG_WF_NUM_COAX}
\end{figure}

\subsection{Minding surface: numerical results}

For the Minding surface we follow the same procedure as before; for the numerics we use $\gamma=1$\,nm and the same radius of curvature $a=60$\,nm as before. As the embeddable part of the Minding surface is finite, the discretization lattice samples the whole surface $\Sigma$. In case of the Minding surface, numerical diagonalization of the Dirac eigenvalue problem corresponding to the Hamiltonian $\widetilde{H}$ with magnetic field $\vec{B}_\perp$ yields the eigenenergies shown in Fig.~\ref{FIG_DISP_MINDING}. 
For high magnetic fields the Landau level spectrum agrees with the analytical results obtained for the pseudosphere in the perpendicular field, Eq.~(\ref{EQ_DISP}), as both carry the same magnetic flux assuming equal $a$ and $B$.
Hence, we have strong numerical evidences that Eq.~(\ref{EQ_DISP}) also applies to the Landau level
spectrum for Minding surface $\Sigma$ with $\gamma << a$ in the perpendicular magnetic field.\\

\begin{figure}[tbp]
  \centering
  \includegraphics[width=0.8 \columnwidth]{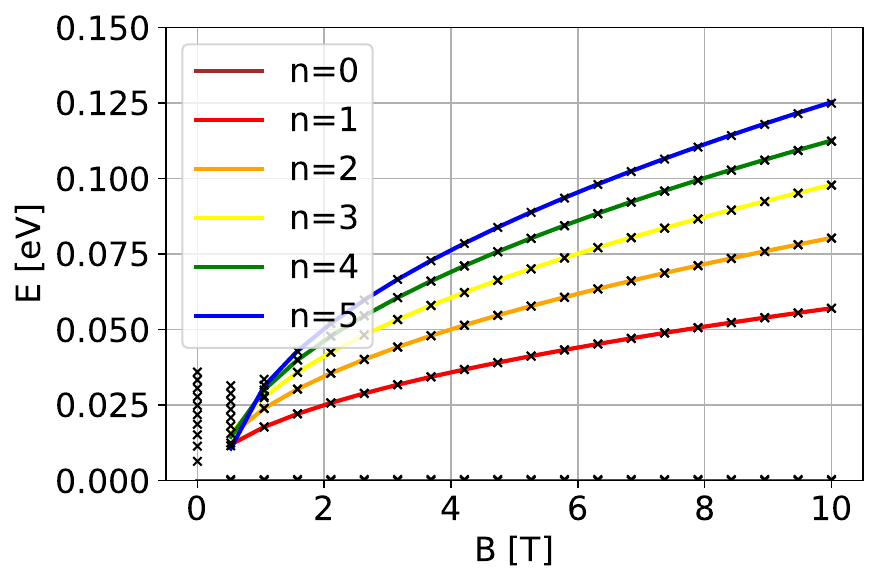}
  \caption{Landau level spectrum of the Minding surface and pseudosphere in a perpendicular magnetic field $\vec{B}_\perp$. Lowest eigenenergies as functions of $B$ are obtained numerically for the Minding surface (black symbols) and analytically for the pseudosphere (solid, colored lines),
  Eq.~(\ref{EQ_DISP}), both with the same radius of curvature $a$.
  Numerical parameters: \(N_W=8\), the arclength lattice constant \(b=0.5\)\,nm, radius of curvature $a=60$\,nm and $\gamma=1$\,nm,
  for the sake of brevity, only zero and positive eigenvalues are shown explicitly.}
  \label{FIG_DISP_MINDING}
\end{figure}


Fig.~\ref{FIG_DISP_NUM_COAX_MINDING} shows the numerical results for a coaxial field and angular quantum numbers $m=-1$ and $m=0$.  
As for the perpendicular field, we complemented the numerical data obtained for the Minding surface with the plots displaying the approximate pseudosphere dispersions $E_{n,m}$, namely, 
Eq.~(\ref{EQ:SpectrumPseudoSphereCoaxial}) for $m=-1$, 
and Eq.~(\ref{EQ:SpectrumPseudoSphereCoaxialMpos}) for $m=0$. As before, matching is remarkably very good.


\begin{figure}[tbp]
  \centering
  \includegraphics[width=0.8\columnwidth]{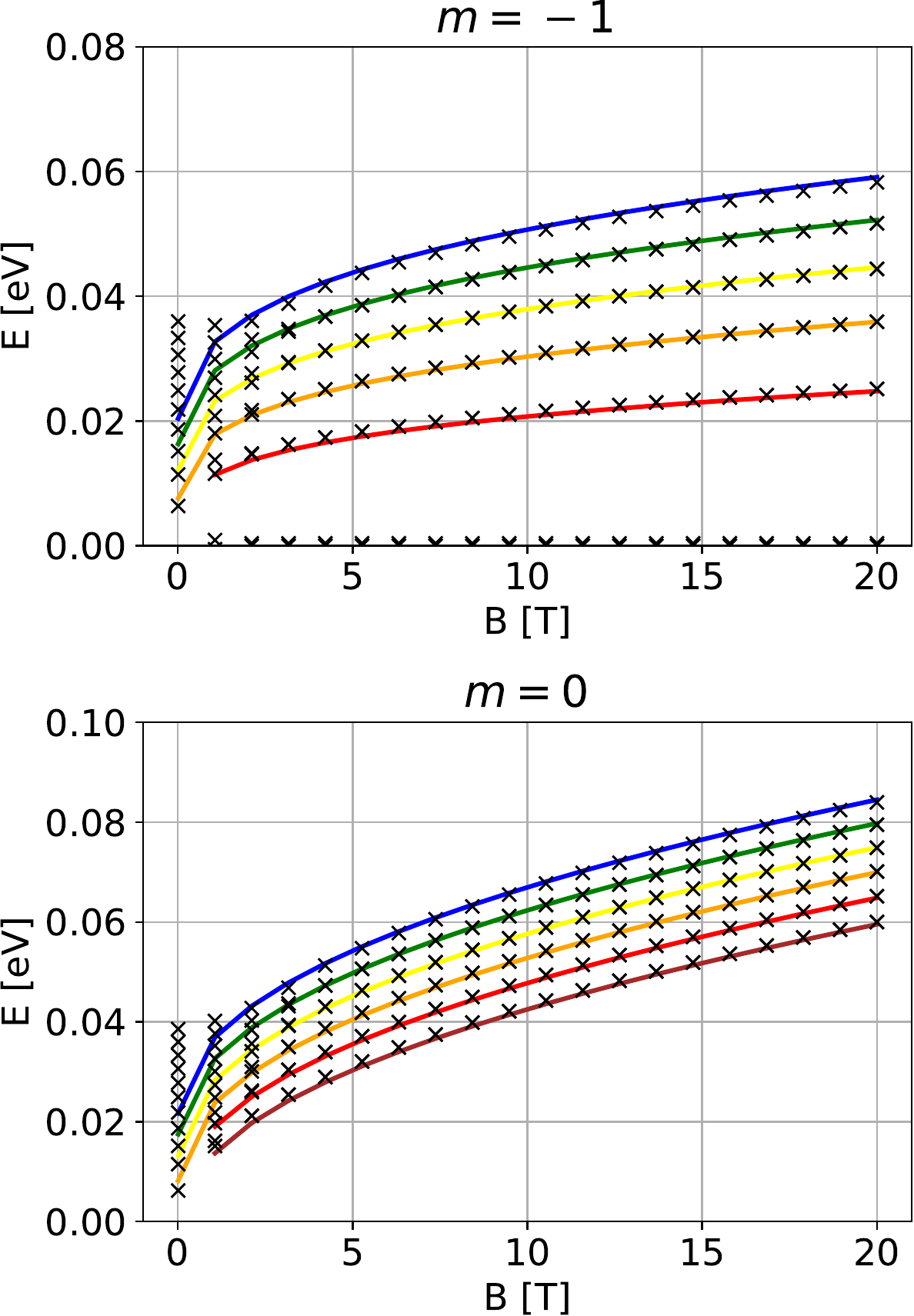}

  \caption{
  Landau level spectrum of the Minding surface in a coaxial field $\vec{B}_\parallel=(0,0,B)$.
  Lowest eigenenergies $E_{n,m}(B)$ (solid, colored lines labeled by principal quantum number) obtained for the case of pseudosphere -- Eq.~(\ref{EQ:SpectrumPseudoSphereCoaxial}) for $m=-1$, and Eq.~(\ref{EQ:SpectrumPseudoSphereCoaxialMpos}) for $m=0$ -- are compared to numerically obtained eigenenergies for the Minding surface (black symbols). 
  Panel~(a) for $m=-1$, panel~(b) for $m=0$.
  Here: Minding surface radius of curvature \(a=60\)\,nm and $\gamma=1$\,nm, the arclength lattice constant \(b=0.5\)\,nm, 
  for the sake of brevity, only zero and positive eigenvalues are displayed.
  }
  \label{FIG_DISP_NUM_COAX_MINDING}
\end{figure}

\section{Conclusion and outlook} \label{SecConclusion}

We analytically and numerically analyze (integer) quantum Hall states of massless Dirac fermions on prominent surfaces of revolution with constant negative Gaussian curvature -- the pseudosphere and the Minding surface -- subjected to perpendicular and coaxial magnetic fields. In special cases -- zeroth Landau levels in $\vec{B}_{\perp/\parallel}$ and higher Landau levels for a pseudosphere in $\vec{B}_{\perp}$ -- we found explicit eigenfunctions.  For all considered situations we provided numerical tight-binding calculations that confirm our analytics with high accuracy.

In the following we compare our results from the spectral perspective and for different geometries, namely, for the flat and positively curved spaces with constant Gaussian curvature.

The Dirac Landau levels on the flat 2D plane---a surface with zero Gaussian 
curvature---in a homogeneous perpendicular field scales as
\begin{equation}
    E_n= \pm C\sqrt{n|B|}\,,\ \ \ \text{where}\ \ n=0,1,2,\dots\,.
\end{equation}
Counting both spin projections, each Landau level is degenerate by means of equal filling factor $|\Phi_\perp|/\Phi_0$, independent of the principal quantum number. Compared to other curved geometries, the zeroth Dirac Landau level in flat space is spin unpolarized.

Similarly, the energies of Dirac Landau levels on a 2D sphere with radius \(a\) in a perpendicular magnetic field (field of magnetic monopole) read
\begin{equation}
E_n = \pm \frac{\hbar v_F}{a} \sqrt{n \left(\frac{|\Phi_\perp|}{\Phi_0}+ n \right)}\,,\ \text{with}\ n=0,1,2,\dots
\end{equation}
where $|\Phi_\perp|/\Phi_0$ again represents the number of flux quanta per a surface of the sphere~\cite{Lee2009,SPHERE} with $|\Phi_\perp|=4\pi a^2 |B|$. Although the Landau levels are specified by the number of flux quanta and the principal quantum number $n$, different Landau levels have different degeneracies, particularly, the $n$-th Landau level for $n\neq 0$ is $(2s_n+1)$ degenerate, where $s_n=\lfloor|\Phi_\perp|/\Phi_0\rfloor+n-1/2$ and $\lfloor\,.\,\rfloor$ stands for the floor function. The zeroth Dirac Landau level is fully spin polarized, with spins parallel to $-\vec{B}_\perp$.

For the pseudosphere with curvature radius $a$ we found both by analytical and numerical calculations that the Dirac Landau levels in a perpendicular field follow
\begin{equation}
E_n = \pm \frac{\hbar v_F}{a} \sqrt{n \left(\frac{|\Phi_\perp|}{\Phi_0} -n \right)} \quad .
\end{equation}
Contrary to flat space and the positively curved sphere, the pseudosphere supports only a finite number of Landau levels with $n$ running from $0$ up to $\lfloor|\Phi_\perp|/\Phi_0\rfloor$, with 
$|\Phi_\perp|=2\pi a^2 |B|$. However, just as in flat space, the degeneracy of each Landau level 
$n$ is given by a filling factor depending on the magnitude of $B$, but is independent of $n$, 
see Eq.~(\ref{Eq:mcritperp}).
Note that the existence of a finite number of Landau levels can be guessed by inspection of the effective mass potential $|V_{\perp/\parallel}(l)|$, see Figs.~\ref{FIG_VPOT} and \ref{FIG_VPOT_COAX}: the potential does not diverge for $l\to0$, yielding a well of finite depth.
Moreover, we have very strong numerical evidences that the above formula also gives the correct Dirac Landau level spectrum for the Minding surfaces. Similarly to the case of a 2D sphere, the zeroth Dirac Landau level on the pseudosphere gets fully spin polarized, though now the spin points along the $+\vec{B}_\perp$ direction, \ie the opposite compared to the 2D sphere.

Apart from the ranges of principal quantum numbers and different degeneracies of the Landau levels, the energy spectrum of massless Dirac electrons on a surface with constant Gaussian curvature $\kappa$ in a perpendicular field $\vec{B}_\perp$ reads in general
\begin{equation}
    E_n=\pm \hbar v_F\sqrt{n(C_\kappa |B|+\kappa n)}\,.
\end{equation}

It is worth to stress that the 2D sphere considered above is closed and compact, as opposed to the 2D plane and pseudosphere, therefore considering only its upper hemisphere 
$C_\kappa$ becomes $\kappa$ independent and equals $2\pi/\Phi_0$.

For the case of the perpendicular magnetic field $\vec{B}_\perp$ we highlight the possibility to spectrally split the radial part of the Hamiltonian, Eq.~(\ref{Eq:RadialDirac}), in terms of $\hat{L}^\pm$ operators. This can be interpreted and approached from the point of view of 1D supersymmetric quantum mechanics \cite{WITTEN1982,Gendenshtein1985,Kochan-Pseudo-2004}. However, we did not follow this approach in our work, but as expected \cite{Gendenshtein1985}, the zeroth Landau level physics is connected to the index theorem of the corresponding Dirac operator.

A further central aspect of our work consists of the study of the coaxial magnetic field configuration, which to the best of our knowledge has hitherto not been considered.  It is however particularly relevant since its experimental realization seems not too far, considering current capabilities in shaping TI nanowires \cite{Ziegler2018, Kessel2017, Behner2023}, which host massless Dirac electrons on their (curved) surfaces.  In such a configuration we did not succeed in finding exact analytical solutions to the problem, but via asymptotic methods we could determine approximate eigenspectra. Our results were fully confirmed by numerical tight-binding calculations.

The spectral structure of the pseudosphere and Minding surface in a coaxial field can be probed, e.g., optically or via transport.  Moreover, it could have interesting implications for orbital magnetic phenomena.  A general question is indeed how the magnetic properties of quantum systems in curved space may be affected by the local (Gaussian) curvature, and whether for example signatures of fractional angular momentum quantization, as proposed in Ref.~\cite{WMAN}, can be identified.


\begin{acknowledgments}
We thank A.~Comtet for a helpful conversation, and S.~Klevtsov and P.~Wiegmann for useful discussions at an early stage of this work. Further, we thank D.-N.~Le for useful comments on their work \cite{Le_2019}, and most importantly we are grateful to M.~Barth for various valuable hints regarding the numerical implementation. CG further acknowledges stimulating discussions with the STherQO members.
We acknowledge support by the
Deutsche Forschungsgemeinschaft (DFG, German Research Foundation) within Project-ID 314695032 -- SFB 1277 (project A07).
D.K.~acknowledges partial support from the project 
IM-2021-26 (SUPERSPIN) funded by the Slovak Academy of Sciences via the programme IMPULZ 2021, and
VEGA Grant No.~2/0156/22---QuaSiModo.
\end{acknowledgments}

\FloatBarrier
\appendix
\section{Spin connection for pedestrians -- adiabatic parallel transport on surfaces of revolution}\label{APP_SPINCON}

2D surfaces of revolution $\Sigma$ embedded in 3D Euclidean space parameterized by $(\phi,r)$ were discussed---including the tangent vectors $\vec{\partial}_\phi$, $\vec{\partial}_r$ and the induced metric tensor $g_{\mu\nu}$---in section \ref{SecDiracCurved}. It is this Euclidean ambient space, 
and the fact that the spinor representations in 2D and 3D spaces are both two-dimensional, 
that allows one to define a spin connection in a very natural manner. 
However, the formal definition as provided by Eq.~(\ref{Eq:SpinConnection}) is intrinsic, i.e.~it is valid for any manifold and is independent of 
any embedding. 

To describe spinor fields on $\Sigma$ we need to specify a quantization axis, that we define point-wisely, i.e.~at any point 
$(\phi,r)$ on $\Sigma$ the local quantization axis is given by the unit vector $\vec{n}$, that corresponds to the outer normal $\vec{n}\propto\vec{\partial}_\phi\times\vec{\partial}_r$ of the surface $\Sigma$ at that point. Here $\times$ is the standard vector product in the 3D ambient Euclidean space. Evaluating $\vec{n}$, using expressions given by Eqs.~(\ref{Eq:TangentVectors}), one gets:
\begin{equation}
  \vec{n}=\frac{1}{\sqrt{1+\left(\frac{dz}{dr}\right)^2}}
  \left(
  \frac{dz}{dr}\cos{\phi},\frac{dz}{dr}\sin{\phi},1
  \right)\,.  
\end{equation}
Since $1+\left(dz/dr\right)^2\geq 1$ one can define an auxiliary angle $\vartheta$ that depends on $r$ and $dz/dr$, 
such that 
\begin{equation}
\cos{\vartheta}=\frac{1}{\sqrt{1+\left(\frac{dz}{dr}\right)^2}}=\hat{\omega}^{12}_\phi\ \Leftrightarrow\  \sin{\vartheta}=\frac{\frac{dz}{dr}}{\sqrt{1+\left(\frac{dz}{dr}\right)^2}}\,,
\end{equation}
or $\tan{\vartheta}=dz/dr$. Doing so one can rewrite the normal vector in a conventional form:
\begin{equation}
    \vec{n}=(\cos{\phi} \sin{\vartheta},\sin{\phi} \sin{\vartheta},\cos{\vartheta})\,.
\end{equation}
Hence, specifying $\vec{n}$ at a given point $(\phi,r)$ means to define point-wisely a spin basis 
that is framed by spin up and spin down states, $\boldsymbol{|+\rangle}_{(\phi,r)}$ and $\boldsymbol{|-\rangle}_{(\phi,r)}$---which are chosen as the eigenstates of $\vec{n}\cdot\vec{\sigma}$. 
Using $\cos{\vartheta}$ and $\sin{\vartheta}$ that depend on $r$ and $dz/dr$, and coordinate $\phi$ 
we can write them in a conventional and very compact way:
\begin{align}
\boldsymbol{|+\rangle}_{(\phi,r)} &=
\left(
\begin{array}{c}
+e^{-i\phi/2}\,\cos{\vartheta/2}\\
\phantom{+}e^{+i\phi/2}\,\sin{\vartheta/2}
\end{array}
\right)\,,\\
\boldsymbol{|-\rangle}_{(\phi,r)} &=
\left(
\begin{array}{c}
-e^{-i\phi/2}\,\sin{\vartheta/2}\\
\phantom{-}e^{+i\phi/2}\,\cos{\vartheta/2}
\end{array}
\right)\,.
\end{align}
This is our particular choice (gauge) of the spin bundle trivialization over the surface $\Sigma$. Someone else can, however, impose a different gauge, i.e.~instead of $|\pm\rangle_{(\phi,r)}$, use 
$e^{if_{\pm}(\phi,r)}\,|\pm\rangle_{(\phi,r)}$ with some locally defined 
phase-functions $f_{\pm}(\phi,r)$ (no global definition over the whole $\Sigma$ is required).
Moreover, observe that we are picking the ``anti-periodic gauge'' i.e.
\begin{equation}
\boldsymbol{|\pm\rangle}_{(\phi+2\pi,r)}=-\boldsymbol{|\pm\rangle}_{(\phi,r)}\,.
\end{equation}
Having a spinor-field $\Psi$ on $\Sigma$, therefore, means to define $\phi$-anti-periodic component fields $\Psi_\ua(\phi,r)$ and $\Psi_\da(\phi,r)$, such that the full spinor
\begin{align}
    \Psi
    &=
    \Psi_\ua(\phi,r)\,\boldsymbol{|+\rangle}_{(\phi,r)}+\Psi_\da(\phi,r)\boldsymbol{|-\rangle}_{(\phi,r)}\nonumber\\
    &\equiv
    \begin{pmatrix}
\Psi_\ua(\phi,r) \\
\Psi_\da(\phi,r)
\end{pmatrix}\,,
\end{align}
is a global object unambiguously defined over the surface $\Sigma$. It is this anti-periodic gauge 
what imposes the ansatz, Eq.~(\ref{EQ_ANGULAR}), taking the angular part of $\Psi$ in the form 
$e^{i(m+{1}/{2})\phi}$.\\ 

The \textit{parallel transport} is a rule prescribing to what spinor will correspond the spinor $\boldsymbol{|\pm\rangle}_{(\phi-d\phi,r-dr)}$ when moved from the given point $(\phi-d\phi,r-dr)$ 
to an infinitesimally close point $(\phi,r)$. Let us denote that parallerly transported spinor at $(\phi,r)$ 
as $\text{PT}\boldsymbol{|\pm\rangle}_{(\phi,r)}$, and let us decompose it with respect to the corresponding $\boldsymbol{|\pm\rangle}_{(\phi,r)}$ states residing at $(\phi,r)$, i.e.
\begin{equation}
 \text{PT}\boldsymbol{|\pm\rangle}_{(\phi,r)}=\mathcal{A}_{\pm}(\phi,r) \boldsymbol{|+\rangle}_{(\phi,r)}+\mathcal{B}_{\pm}(\phi,r) \boldsymbol{|-\rangle}_{(\phi,r)}\,.   
\end{equation}
So technically, the parallel transport is given by a prescription for the position dependent coefficients
$\mathcal{A}_{\pm}$ and $\mathcal{B}_{\pm}$.

From the point of view of 2D surface, the spinors $\boldsymbol{|\pm\rangle}_{(\phi-d\phi,r-dr)}$ and $\boldsymbol{|\pm\rangle}_{(\phi,r)}$, which are residing at different (although nearby) points do 
not know about each other, so something like their sum, or inner product do not have a priori 
any sense. However, from the point of view of the ambient 3D Euclidean space and its flat spin structure one can move $\boldsymbol{|\pm\rangle}_{(\phi-d\phi,r-dr)}$ and $\boldsymbol{|\pm\rangle}_{(\phi,r)}$ to the origin of $\mathbf{R}^3$ and perform their scalar product there etc. 
The rule we prescribe for $\text{PT}\boldsymbol{|\pm\rangle}$---i.e.~rules for $\mathcal{A}_{\pm}$ and $\mathcal{B}_{\pm}$---relies on that ambient space, namely, we define
\begin{align}
    \mathcal{A}_{+}&=\langle\langle +_{(\phi,r)}|+_{(\phi-d\phi,r-dr)}\rangle\rangle\simeq 1-i\tfrac{d\phi}{2}\cos{\vartheta}\,,\\
   \mathcal{A}_{-}&=\langle\langle +_{(\phi,r)}|-_{(\phi-d\phi,r-dr)}\rangle\rangle\simeq 0\,,\\
    \mathcal{B}_{+}&=\langle\langle -_{(\phi,r)}|+_{(\phi-d\phi,r-dr)}\rangle\rangle\simeq 0\,,\\
    \mathcal{B}_{-}&=\langle\langle -_{(\phi,r)}|-_{(\phi-d\phi,r-dr)}\rangle\rangle\simeq 1+i\tfrac{d\phi}{2}\cos{\vartheta}\,,
\end{align}
where the meaning of $\simeq$ on the right-hand sides means an expansion up to the first order in $d\phi$ and $d\vartheta\propto dr$. More importantly, the scalar product 
$\langle\langle\,\cdot\, |\,\cdot\, \rangle\rangle$ entering the above definition is understood in 
a sense of the standard two-component $\mathbf{C}^2$ Pauli spinors in the ambient Euclidean space $\mathbf{R}^3$, and not in a sense of Eq.~(\ref{Eq:ScalarProductGeneral}). Since the parallel transport
preserves spin projections on infinitesimal distances, $\mathcal{A}_{-}=0=\mathcal{B}_{+}$, and preserves norms, $|\mathcal{A}_{+}|=1=|\mathcal{B}_{-}|$, 
both up to the first order in $d\phi$ and $dr$, it is called the \textit{spin-adiabatic parallel-transport} or \textit{spin connection}, but in an essence it is the Berry connection.

Covariant derivatives, or more precisely, their spinor part components---see Eqs.~(\ref{Eq:D1}) and (\ref{Eq:D2})---are defined as:
\begin{align}
\nabla_r \boldsymbol{|\pm\rangle}_{(\phi,r)}
&:=
\lim\limits_{dr\rightarrow 0}
\frac{\boldsymbol{|\pm\rangle}_{(\phi,r)}-\text{PT}\boldsymbol{|\pm\rangle}_{(\phi,r)}}{d r} \notag \\
&=0\,,\\
\nabla_\phi \boldsymbol{|\pm\rangle}_{(\phi,r)}
&:=
\lim\limits_{d\phi\rightarrow 0}
\frac{\boldsymbol{|\pm\rangle}_{(\phi,r)}-\text{PT}\boldsymbol{|\pm\rangle}_{(\phi,r)}}{d\phi}\notag \\
&=\pm\frac{i}{2}\cos{\vartheta}\,\boldsymbol{|\pm\rangle}_{(\phi,r)}\notag\\
&=\frac{i}{2} \hat{\omega}_\phi^{12} \sigma_z\,\boldsymbol{|\pm\rangle}_{(\phi,r)}\,,
\end{align}
what is exactly the spin connection we have encountered in section~\ref{SecDiracCurved}. So $\nabla_r$ and $\nabla_\phi$
``quantify'' how much the spin-bundle trivialization, i.e.~states $\boldsymbol{|\pm\rangle}_{(\phi,r)}$, changes when transported infinitesimally between neighbouring points on $\Sigma$.


\section{Christoffel symbols}\label{APP_CHRISTOFFEL}

An elegant way to obtain the Christoffel symbols for a given metric and curved coordinates is to use the Lagrange function---kinetic energy of a unit mass particle on the curved surface \cite{D},
\begin{equation}
\mathcal{L}(\phi, \dot{\phi}, r, \dot{r}) = \frac{1}{2}(\dot{\phi},\dot{r})
\begin{pmatrix}
   r^2 & 0 \\ 0 & 1 + \left(\frac{dz(r)}{dr}\right)^2
\end{pmatrix}
\begin{pmatrix}
   \dot{\phi} \\ \dot{r}
\end{pmatrix}\,,
\end{equation}
where the matrix represents the spatial part of the space-time metric tensor 
$g_{\mu\nu}=\text{diag}(-v_F^2,r^2,1 + \left(dz(r)/dr\right)^2)$ 
and the dot denotes the derivative with respect to a curve parameter, say time, on which both coordinates, 
\(\phi\) and \(r\) depend upon. 
Writing down the corresponding Euler-Lagrange equation for \(\phi\) one recovers the Christoffel symbols 
$\Gamma^{\phi}_{..}$ \cite{D},
\begin{equation}
\frac{d}{dt} \left(\frac{\partial \mathcal{L}}{\partial \dot{\phi}}\right) - \frac{\partial \mathcal{L}}{\partial \phi} = 2 r \dot{r} \dot{\phi} + r^2 \ddot{\phi} = 0 \,.
\end{equation}
This is equivalent to
\begin{equation}
	2  \frac{1}{r} \dot{r} \dot{\phi} + \ddot{\phi} \overset{!}{=} \Gamma^\phi_{r \phi} \dot{r} \dot{\phi} + \Gamma^\phi_{\phi r} \dot{\phi} \dot{r} + \ddot{\phi} = 0 \,,
\end{equation}
out of which the non-zero Christoffel symbols read
\begin{equation}\label{CHST}
\Gamma^{\phi}_{\phi r} = \Gamma^\phi_{r \phi} = \frac{1}{r} \,.
\end{equation}
Repeating the same for the $r$ coordinate, one recovers the corresponding non-zero $\Gamma^{r}_{..}$ Christoffel symbols:
\begin{equation}
    \Gamma^{r}_{rr} = \frac{\frac{dz}{dr} \frac{d^2z}{dr^2}}{1+\left(\frac{dz}{dr} \right)^2} \quad \Gamma^{r}_{\phi \phi} = - \frac{r}{1+\left(\frac{dz}{dr} \right)^2}
\end{equation}

\section{Solutions of the radial equations}\label{APP_RADIAL}
The differential equation we need to solve for the $E\neq 0$ states in perpendicular magnetic field, see Eq.~(\ref{Eq:AnsatzForHigherLLPolynomial}), is of the form
\begin{equation}
x^2 \frac{d^2y(x)}{dx^2} + (\alpha x +\beta) \frac{dy(x)}{dx} + \gamma y(x) = 0 \,,
\end{equation}
and its coefficients $\alpha$, $\beta$ and $\gamma$ follow from 
Eq.~(\ref{Eq:AnsatzForHigherLLPolynomial}). This is a \emph{degenerate double-confluent Heun equation} which can be simplified using the transformation \cite{Figueiredo_2005}
\begin{equation}
\Tilde{x} = \tfrac{\beta}{x}, \, y(\Tilde{x}) = \Tilde{x}^{\Delta_{\pm}} \Tilde{y}(\Tilde{x}), \, \Delta_\pm = \tfrac{\alpha -1 \pm \sqrt{\left(\alpha -1 \right)^2-4 \gamma}}{2} \, .
\end{equation}
This leads to a \emph{confluent hypergeometric equation}
\begin{equation}
\Tilde{x} \frac{d^2 \Tilde{y}(\Tilde{x})}{d \Tilde{x}^2} + \left[\left(2 \Delta_\pm + 2 - \alpha \right) - \Tilde{x} \right] \frac{d \Tilde{y}(\Tilde{x})}{d \Tilde{x}} - \Delta_\pm \Tilde{y}(\Tilde{x}) = 0\,.
\end{equation}
For a given $\Delta$, this equation has two linearly independent solutions, for instance \emph{Kummer's function 
$M(a, b, z)$} and \emph{Tricomi's function $U(a, b, z)$}. 
Counting the possibility to choose $\Delta_+$ or $\Delta_-$ in the above transformation, 
this would seemingly yield four solutions. Of course, not all of them can be linearly independent since the original equation is only of second order. A basis of solutions may be given by
\begin{equation}
y_-(x) = x^{-\Delta_-} M \left(\Delta_-, 2 \Delta_- +2 - \alpha, \frac{\beta}{x}\right) \, ,
\end{equation}
\begin{equation}
y_+(x) = x^{-\Delta_+} M \left(\Delta_+, 2 \Delta_+ +2 - \alpha, \frac{\beta}{x}\right) \, .
\end{equation}
Both solutions become polynomials in case the first argument of $M$ becomes a non-positive integer. Before we examine this further, we want to show that for our problem only polynomial solutions lead to physical, i.e.~normalizable, wavefunctions. For this we need to inspect the asymptotic behavior of the whole wavefunction in the limit $\Tilde{r} \to 0$ and hence in particular $M(a, b, z)$ for $z \to \infty$ and $z \to -\infty$, assuming that the first parameter is not a non-positive integer.  The first limit, $z \to \infty$, is given by
\begin{equation}
z \to \infty: \;M(a, b, z) \to \frac{e^z z^{a-b}}{\Gamma(a)} \, .
\end{equation}
For $z \to -\infty$, we use \emph{Kummer's transformation}
\begin{equation}
M(a, b, z) = e^z M(b-a, b, -z) 
\end{equation}
which gives us the relation
\begin{equation}
z \to -\infty: \; M(a, b, z) = e^{-z} M(b-a, b, -z) \to \frac{z^{-a}}{\Gamma(b-a)} \, .
\end{equation}
Recall that in our original ansatz for $\Psi^{\uparrow }_{\perp}(\Tilde{r})$, Eq.~(\ref{Eq:AnsatzForHigherLL}), we used a function which in the present context has 
asymptotic behavior $e^{-\frac{\beta}{2x}}$. 
So assuming $\beta$ is positive (we need the limit $z \to \infty$), the overall behavior is $e^{+\frac{\beta}{2x}}$, which clearly blows up when $x \to 0$. In case of a negative $\beta$ (we need the limit $z \to - \infty$), the asymptotic behavior is given by $e^{-\frac{\beta}{2x}}$, which also blows up. This clearly shows that only polynomial solutions yield physical wavefunctions.
The condition that at least one of the two basis solutions behaves polynomially is given by
\begin{equation}
\Delta_\pm \overset{!}{=} -n, \, n \in \{0, 1, 2, \dots \} \,,
\end{equation}
what gives us the quantization condition for the eigenenergies. 
For given $\alpha$ and $\beta$, it turns out that this equation can only be solved for $\Delta_-$. The solution reads
\begin{equation}
\gamma = -n^2 -n \left(\alpha -1 \right) \, .
\end{equation}
Overall, our solution for the radial differential equation reads
\begin{equation}
y(x) = x^n M \left(-n, -2n +2 - \alpha , \frac{\beta}{x} \right) \, .
\end{equation}
This can be simplified even further employing the \emph{associated Laguerre polynomials $L_n^{(a)}(x)$}
\begin{equation}
y(x) = x^n L_n^{(-2n+1-\alpha)} \left(\frac{\beta}{x} \right) \, ,
\end{equation}
where we have absorbed some constants into the overall normalization (not shown).\\

The differential equations needed for the coaxial magnetic field, Eq.~(\ref{Eq:AnsatzForHigherLLPolynomialCoax}), can be brought into the form
\begin{equation}
x^2 \frac{d^2y(x)}{dx^2} + (\alpha x^2 + x + \beta) \frac{dy(x)}{dx} + \gamma y(x) = 0 \,,
\end{equation}
respectively
\begin{equation}
x^2 \frac{d^2y(x)}{dx^2} + (\alpha x^2 + x + \beta) \frac{dy(x)}{dx} + \left(\gamma + \alpha x \right) y(x) = 0 \,,
\end{equation}
with the coefficients $\alpha$, $\beta$ and $\gamma$ stemming from Eq.~(\ref{Eq:AnsatzForHigherLLPolynomialCoax}). 
Those are \emph{double-confluent Heun equations}. Contrary to the previous case, they are not degenerate (for $\alpha \neq 0$, $\beta \neq 0$) and hence a transformation into the much simpler confluent hypergeometric differential equation is not possible. 
However, further insight can be gained when performing the rescaling
\begin{equation}
\Tilde{x} = -\alpha x \,,
\end{equation}
which yields
\begin{equation}
\Tilde{x}^2 \frac{d^2y(\Tilde{x})}{d\Tilde{x}^2} + (- \Tilde{x}^2 + \Tilde{x} - \alpha \beta) \frac{dy(\Tilde{x})}{d\Tilde{x}} + \gamma y(\Tilde{x}) = 0 \,,
\end{equation}
respectively
\begin{equation}
\Tilde{x}^2 \frac{d^2y(\Tilde{x})}{d\Tilde{x}^2} + (- \Tilde{x}^2 + \Tilde{x} - \alpha \beta) \frac{dy(\Tilde{x})}{d\Tilde{x}} + \left(\gamma - \Tilde{x} \right) y(\Tilde{x}) = 0 \,.
\end{equation}
Interestingly, the coefficients $\alpha$ and $\beta$ do not enter independently into the equation. As in our original problem $\gamma$ plays a role of the energy eigenvalue, the above equation then asserts that $\gamma$ depends only on the product $\alpha \beta$ and not on their separate values. 
The solutions of the differential equation can be expressed in terms of the \emph{double-confluent Heun function}. Since it is quite involved to work with them, we relax an attempt to provide the wavefunctions in analytic form and rather focus on the eigenenergies. To our best knowledge, there is currently no exact formula determining the spectrum of this equation. However, there are known asymptotic methods and approximations \cite{W_Lay_1998} which allow to get the eigenvalues expanded in reciprocal powers of $\sqrt{- \alpha \beta}$, in particular up to second order,
\begin{equation}
\gamma_n = 2n \sqrt{-\alpha \beta} + \frac{1}{2} n^2 -1\,,
\end{equation}
respectively
\begin{equation}
\gamma_n = (2n+1) \sqrt{-\alpha \beta} + \frac{1}{2} n^2 + \frac{1}{2} n - 2\,.
\end{equation}
Using this formula and expressing $\alpha$ and $\beta$ in terms of the original parameters, one recovers the asymptotic eigenvalue results as given by Eq.~(\ref{EQ:SpectrumPseudoSphereCoaxial}).

\section{WKB analysis for the pseudosphere in a coaxial field} 
\label{SEC_WKB}
In the following we perform a WKB analysis of the spectrum of the pseudosphere in a coaxial field  starting from Eq.~(\ref{Eq:FirstAndSecond3}). It turns out to be advantageous to work with the arclength coordinate $l$. The Hamiltonian is then given by
\begin{eqnarray}
\hat{H} &=& -\partial_l^2 + (\partial_l V) + V^2 \, ,
\end{eqnarray}
with the effective potential
\begin{eqnarray}
V = \frac{m+\frac{1}{2}}{a} \exp \left(\frac{l}{a} \right) + \frac{1}{2 a \Tilde{l}_B^2} \exp \left(-\frac{l}{a} \right) \, .
\end{eqnarray}
Eigenvalues of $\hat{H}$ are then $\epsilon^2 := \frac{E^2}{\hbar^2 v_F^2}$. As a first approximation, we can neglect the derivative of $V$ in $\hat{H}$
\begin{eqnarray}
\hat{H} &\approx& -\partial_l^2 + V^2 \\
&=& -\partial_l^2 + \frac{\left(m + \frac{1}{2} \right)^2}{a^2} \exp \left(\frac{2l}{a} \right) + \frac{1}{4 a^2 \Tilde{l}_B^4} \exp \left(-\frac{2l}{a} \right) \notag \\
&& + \frac{m+\frac{1}{2}}{a^2 \Tilde{l}_B^2} \, . \notag
\end{eqnarray}
We then absorb a constant represented by the last term as a shift of the origin of energy for the states with a given magnetic number $m$, therefore:
\begin{eqnarray}
\hat{\Tilde{H}} &=& -\partial_l^2 \\
&&+ \frac{\left(m + \frac{1}{2} \right)^2}{a^2} \exp \left(\frac{2l}{a} \right) + \frac{1}{4 a^2 \Tilde{l}_B^4} \exp \left(-\frac{2l}{a} \right) \, . \notag
\end{eqnarray}
Performing a shift of the arclength coordinate such that its origin coincides with the minimum of $V^2$ simplifies this even more yielding
\begin{eqnarray}
\hat{\Tilde{H}} &=& - \partial_l^2 + \frac{\eta}{a^2} \cosh \left(\frac{2l}{a} \right) \, ,
\end{eqnarray}
where we introduced $\eta = \left| \frac{m + \frac{1}{2} }{\Tilde{l}_B^2} \right|$. 
The classical momentum is given by
\begin{eqnarray}
p(\Tilde{\epsilon}) &=& \hbar \sqrt{\Tilde{\epsilon}^2 - \frac{\eta}{a^2} \cosh \left(\frac{2l}{a} \right)} \, .
\end{eqnarray}
For a given (shifted) energy $\Tilde{\epsilon}$, we assume the two turning points $l_-, l_+$ to be located within the pseudosphere (obviously $l_- = - l_+$). 
We need to calculate the action of a closed orbit $\gamma$
\begin{eqnarray}
I(\Tilde{\epsilon}) &=& \oint_{\gamma} dl \, p(\Tilde{\epsilon}) \\
&=& 4 \int_0^{l_+} dl \, p(\Tilde{\epsilon}) \notag \\
&=& -i 4 a \hbar \Tilde{\epsilon} \sqrt{1-\frac{\eta^2}{a^2 \Tilde{\epsilon}^2}} \, \text{E} \left(\frac{i}{2} \text{arcosh} \left(\frac{a^2 \Tilde{\epsilon}^2}{\eta} \right), \frac{2 \eta}{\eta - a^2 \Tilde{\epsilon}^2} \right) \notag \\
&\approx& 4 a \hbar \Tilde{\epsilon} \ln \left(\frac{a^2 \Tilde{\epsilon}^2}{\eta} \right) \notag \, ,
\end{eqnarray}
where $\text{E}(\phi, k)$ is the \emph{incomplete elliptic integral of the second kind}.
The particle motion includes two soft turning points at $l_+$ and $l_-$, yielding the corresponding WKB quantization condition
\begin{eqnarray}
\frac{I(\Tilde{\epsilon})}{2 \pi \hbar} \overset{!}{=} n + \frac{1}{2} \, .
\end{eqnarray}
Solving for $\Tilde{\epsilon}$ and adding the energy shift yields the WKB spectrum (normalized to $\mathcal{E}=\hbar v_F/a$)
\begin{eqnarray}
E_{nm} &=& \sqrt{\frac{\pi^2 \left(n + \frac{1}{2} \right)^2}{4\text{W}^2 \left(\frac{\pi}{2} \left(n + \frac{1}{2} \right) \eta^{-1/2} \right)} + \text{sgn} \left(m + \frac{1}{2} \right) \eta} \, , \notag \\
&&
\end{eqnarray}
where $\text{W}(x)$ is the \emph{Lambert W function}.

To examine the behaviour in the strong field limit we assume
\begin{eqnarray}
\frac{1}{\text{W}^2(x)} &\approx& \left(\frac{1}{x-x^2}\right)^2 \\
&=& \left(\frac{1}{x} + \frac{1}{1-x}\right)^2 \notag \\
&\approx& \left(\frac{1}{x} + 1\right)^2 \notag \\
&\approx& \frac{1}{x^2} + \frac{2}{x} \notag \, .
\end{eqnarray}
This yields
\begin{eqnarray} \label{DISP_WKB}
E_{n, m} &\approx& \sqrt{\eta + \text{sgn} \left(m + \frac{1}{2} \right) \eta + \pi \left(n + \frac{1}{2} \right) \sqrt{\eta}}  \notag \, . \\
&& 
\end{eqnarray}
This explains where the different asymptotic behaviour of the spectrum for negative and non-negative angular quantum numbers originates. For the first mentioned, the first two terms under the square root cancel and the third term determines the strong field behaviour $E_{n, m} \propto |B|^{1/4}$. In the  case of non-negative $m$, the first two terms add and are dominant, leading to $E_{n, m} \propto |B|^{1/2}$.

\bibliography{lit.bib}

\end{document}